\documentstyle[prd,eqsecnum,aps,preprint,tighten,amssymb]{revtex}

\def\gsim{\mathrel{%
\rlap{\raise 0.511ex \hbox{$>$}}{\lower 0.511ex
\hbox{$\sim$}}}}

\def\lsim{\mathrel{
\rlap{\raise 0.511ex \hbox{$<$}}{\lower 0.511ex
\hbox{$\sim$}}}}

\begin{document}

\title{Self-gravitating fundamental strings and black holes}
\author{Thibault Damour}
\address{Institut des Hautes Etudes Scientifiques, 91440
Bures-sur-Yvette, France}
\author{Gabriele Veneziano}
\address{Theory Division, CERN, CH-1211 Geneva 23, Switzerland}

\maketitle

\begin{abstract}
The configuration of typical highly excited $(M \gg M_s \sim
(\alpha')^{-1/2})$ string states is considered as the string 
coupling $g$ is adiabatically increased. The size distribution  of 
very massive single string states is studied and the mass shift, due 
to long-range gravitational, dilatonic and axionic attraction, is
estimated. By combining the two effects, in any number of spatial 
dimensions $d$, the most probable size of a string state becomes of 
order $\ell_s = \sqrt{2 \, \alpha'}$ when $g^2 M / M_s \sim1$. 
Depending on the dimension $d$, the transition between a
random-walk-size string state (for low $g$) and a compact ($\sim
\ell_s$) string state (when $g^2 M / M_s \sim 1$) can be very 
gradual ($d=3$), fast but continuous ($d=4$), or discontinuous ($d 
\geq 5$). Those compact string states look like nuggets of an 
ultradense state of string matter, with energy density $\rho \sim 
g^{-2} M_s^{d+1}$. Our results extend and clarify previous work by 
Susskind, and by Horowitz and Polchinski, on the correspondence 
between self-gravitating string states and black holes.

\end{abstract}
\newpage
\section{Introduction}
Almost exactly thirty years ago the study of the spectrum
of string theory (known at the time as the dual resonance model)
revealed \cite{FVBM} a huge degeneracy of states growing as an 
exponential of the mass. This led to the concept of a limiting 
(Hagedorn) temperature $T_{\rm Hag}$ in string theory.
Only slightly more recently Bekenstein \cite{Bekenstein} proposed
that the entropy of a black hole should be proportional to the area 
of its horizon in Planck units, and Hawking \cite{Hawking} fixed the 
constant of proportionality after discovering that black holes, 
after all, do emit thermal radiation at a temperature $T_{\rm Haw} 
\sim R_{\rm BH}^{-1}$.

When string and black hole entropies are compared one immediately
notices a striking difference: string entropy\footnote{The 
self-interaction of a string lifts the huge degeneracy of free string 
states. One then defines the entropy of a narrow band of string 
states, defined with some energy resolution $M_s \lsim \Delta \, E 
\ll M$, as the logarithm of the number of states within the band 
$\Delta \, E$.} is proportional to the first power of mass in any 
number of spatial dimensions $d$, while black hole entropy is 
proportional to a $d$-dependent power of the mass, always larger 
than $1$. In formulae:
\begin{equation}
S_s \sim {\alpha' M \over \ell_s} \sim M / M_s ~~~~ , ~~~~~~
 S_{\rm BH} \sim \frac{{\rm Area}}{G_N} \sim \frac{R_{\rm
BH}^{d-1}}{G_N} \sim \frac{(g^2 \, M / 
M_s)^{\frac{d-1}{d-2}}}{g^2}\; ,
\label{entropies}
\end{equation}
where, as usual, $\alpha'$ is the inverse of the classical string
tension, $\ell_s \sim \sqrt{\alpha' \hbar}$ is the quantum length 
associated with it\footnote{Below, we shall use the precise 
definition $\ell_s \equiv \sqrt{2 \alpha' \hbar}$, but, in this 
section, we neglect factors of order unity.}, $M_s \sim \sqrt{\hbar 
/ \alpha'}$ is the corresponding string mass scale, $R_{\rm BH}$ is 
the Schwarzschild radius associated with $M$:
\begin{equation}
R_{\rm BH} \sim (G_N \, M)^{1/(d-2)} \; ,\label{eq1.1}
\end{equation}
and we have used that, at least at sufficiently small coupling, the 
Newton constant and $\alpha'$ are related via the string coupling by
$G_N \sim g^2 (\alpha')^{(d-1)/2}$ (more geometrically, 
$\ell_P^{d-1}\sim g^2 \ell_s^{d-1}$).

Given their different mass dependence, it is obvious that, for a 
given set of the  fundamental constants $G_N, \alpha', g^2$,
$S_s > S_{\rm BH}$ at sufficiently small $M$, while the opposite is 
true at sufficiently large $M$. Obviously, there has to be a 
critical value of $M$, $M_c$, at which $S_s = S_{\rm BH}$. This 
observation led Bowick et al. \cite{Bowick} to conjecture that large 
black holes end up their Hawking-evaporation process when $M = M_c$, 
and then transform into a higher-entropy string state without ever 
reaching the singular zero-mass limit. This reasoning is confirmed  
\cite{GVFC} by the observation that, in string theory, the 
fundamental string length $\ell_s$ should set a minimal value for 
the Schwarzschild radius of any black hole (and thus a maximal value 
for its Hawking temperature). It was also noticed \cite{Bowick}, 
\cite {susskind}, \cite{GVDivonne} that, precisely at $M= M_c$, 
$R_{\rm BH} = \ell_s$ and the Hawking temperature equals the 
Hagedorn temperature of string theory. For any $d$, the value of 
$M_c$ is given  by:
\begin{equation}
M_c \, \sim M_s g^{-2} \, . \label{eq1.2}
\end{equation}

Susskind and collaborators \cite 
{susskind}, \cite{halyo} went a step further and proposed that the 
spectrum of black holes and the 
spectrum of single string states be ``identical'', in the sense 
that there be a one to one correspondence between (uncharged) 
fundamental string states and (uncharged) black hole states. Such a 
``correspondence principle'' has been generalized by Horowitz and 
Polchinski \cite{hp1} to a wide range of charged black hole states 
(in any dimension). Instead of keeping fixed the fundamental 
constants and letting $M$ evolve by evaporation, as considered 
above, one can (equivalently) describe the physics of this 
conjectured correspondence by following a narrow band of states, on 
both sides of and through, the string $\rightleftharpoons$ black 
hole transition, by keeping fixed the entropy\footnote{One uses here 
the fact that, during an adiabatic variation of $g$, the entropy of 
the black hole $S_{\rm BH} \sim ({\rm Area}) / G_N \sim R_{\rm 
BH}^{d-1} / G_N$ stays constant. This result (known to hold in the 
Einstein conformal frame) applies also in string units because 
$S_{\rm BH}$ is dimensionless.} $S = S_s = S_{\rm BH}$, while 
adiabatically\footnote{The variation of $g$ can be seen, depending 
on one's taste, either as a real, adiabatic change of $g$ due to a 
varying dilaton background, or as a mathematical way of following 
energy states.} varying the string coupling $g$, i.e. the ratio 
between $\ell_P$ and $\ell_s$. The correspondence principle then 
means that if one increases $g$ each (quantum) string state should 
turn into a (quantum) black hole state at sufficiently strong 
coupling, while, conversely, if $g$ is decreased, each black hole 
state should ``decollapse'' and transform into a string state at 
sufficiently weak coupling. For all the reasons mentioned above, it 
is very natural to expect that, when starting from a black hole state, 
the critical value of $g$ at which a 
black hole should turn into a string is given, in clear relation to
(\ref{eq1.2}), by
\begin{equation}
g_c^2 \, M \sim M_s \, , \label{eq1.2'}
\end{equation}
and is related to the common value of string and black-hole entropy 
via
\begin{equation}
g_c^2 \sim \frac{1}{S_{\rm BH}} = \frac{1}{S_s}\; . \label{eq1.2''}
\end{equation}
Note that $g_c^2 \ll 1$ for the very massive states ($M \gg
M_s$) that we consider. This justifies our use of the perturbative 
relation between $G_N$ and $\alpha'$.

In the case of extremal BPS, and nearly extremal, black holes
the conjectured correspondence was dramatically confirmed
through the work of  Strominger and Vafa \cite{SV} and others
\cite{others} leading to a statistical mechanics interpretation of
black-hole entropy in terms of the number of microscopic states 
sharing the same macroscopic quantum numbers. However, little is 
known about whether and how the correspondence works for 
non-extremal, non BPS black holes, such as the simplest  
Schwarzschild black hole\footnote{For simplicity, we shall
consider in this work only Schwarzschild black holes, in any number 
$d \equiv D-1$ of non-compact spatial dimensions.}. By contrast to BPS 
states whose mass is protected by supersymmetry, we shall consider here 
the effect of varying $g$ on the mass and size of non-BPS string states. 

Although it is remarkable that black-hole and string entropy
coincide when $R_{\rm BH} = \ell_s$, this is still not quite
sufficient to claim that, when starting from a string state, a string 
becomes a black hole at $g = g_c$.
In fact, the process in which one starts from a string state in flat 
space and increases $g$ poses a serious puzzle \cite{susskind}. 
Indeed, the radius of a typical excited string state of mass $M$ is 
generally thought of being of order
\begin{equation}
R_s^{\rm rw} \sim \ell_s (M / M_s)^{1/2} \, , \label{eq1.5}
\end{equation}
as if a highly excited string state were a random walk made of
$M/M_s = \alpha'M/\ell_s$ segments of length $\ell_s$ \cite{rw}. 
[The number of steps in this random walk is, as is natural, the 
string entropy (\ref{entropies}).] The ``random walk'' radius 
(\ref{eq1.5}) is much larger than the Schwarzschild radius for all 
couplings $g \le g_c$, or, equivalently, the ratio of 
self-gravitational binding energy to mass (in $d$ spatial 
dimensions)
\begin{equation}
\frac{G_N \, M}{(R_s^{\rm rw})^{d-2}} \sim \left( \frac{R_{\rm BH}
(M)}{R_s^{\rm rw}} \right)^{d-2} \sim g^2 \left( \frac{M}{M_s}
\right)^{\frac{4-d}{2}} \label{eq1.6}
\end{equation}
remains much smaller than one (when $d>2$, to which we restrict
ourselves) up to the transition point. In view of (\ref{eq1.6}) it
does not seem natural to expect that a string state will 
``collapse'' to a black hole when $g$ reaches the value 
(\ref{eq1.2'}). One would expect a string state of mass $M$ to turn 
into a black hole only when its typical size is of order of $R_{\rm 
BH} (M)$ (which is of order $\ell_s$ at the expected transition 
point (\ref{eq1.2'})). According to Eq.~(\ref{eq1.6}), this seems to 
happen for a value of $g$ much larger than $g_c$.

Horowitz and Polchinski \cite{hp2} have addressed this puzzle by
means of a ``thermal scalar'' formalism \cite{chi}. Their results
suggest a resolution of the puzzle when $d=3$ (four-dimensional
spacetime), but lead to a rather complicated behaviour when $d \geq
4$. More specifically, they consider the effective field theory of a
complex scalar field $\chi$ in $d$ (spacetime) dimensions (with
period $\beta$ in Euclidean time $\tau$), with mass squared $m^2
(\beta) = (4\pi^2 \, \alpha'^2)^{-1} \, [G_{\tau \tau} \, \beta^2 -
\beta_{\rm H}^2]$, where $G_{\mu \nu}$ is the string metric and
$\beta_{\rm H}^{-1}$ the Hagedorn temperature. They took into 
account the effect of gravitational (and dilatonic) 
self-interactions in a mean field approximation. This leads to an 
approximate Hartree-like equation for $\chi (\mbox{\boldmath$x$})$, 
which admits a stable bound state, in some range $g_0 < g < g_c$, 
when $d=3$. They interpret the size of the bound state ``wave 
function'' $\chi (\mbox{\boldmath$x$})$ as the ``size of the 
string'', and find that (in $d=3$) this size is of order
\begin{equation}
\ell_{\chi} \sim \frac{1}{g^2 \, M} \, . \label{eq1.6bis}
\end{equation}

They describe their result by saying that ``the string contracts 
from its initial (large) size'', when $g \sim g_0 \sim 
(M/M_s)^{-3/4}$, down to the string scale when $g \sim g_c \sim 
(M/M_s)^{-1/2}$. This interpretation of the length scale 
$\ell_{\chi}$, characterizing the thermal scalar bound state, as 
``the size of the string'' is unclear to us, because of the formal 
nature of the auxiliary field $\chi$ which has no direct physical 
meaning in Minkowski spacetime. Moreover, the analysis of 
Ref.~\cite{hp2} in higher dimensions is somewhat inconclusive, and 
suggests that a phenomenon of hysteresis takes place (when $d \geq 
5$): the critical value of $g$ corresponding to the string 
$\rightleftharpoons$ black hole transition would be $g_0 \sim 
(M/M_s)^{(d-6)/4} > g_c$ for the direct process (string 
$\rightarrow$black hole), and $g_c$ for the reverse one. Finally, 
they suggest that, in the reverse process, a black hole becomes an 
excited string in an {\it atypical} state.

The aim of the present work is to clarify the string 
$\rightleftharpoons$ black hole transition by a direct study, in 
real spacetime, of the size and mass of a {\it typical} excited 
string, within the microcanonical ensemble of {\it self-gravitating} 
strings. Our results lead to a rather simple picture of the 
transition, in any dimension. We find no hysteresis
phenomenon in higher dimensions. The critical value for the 
transition is (\ref{eq1.2'}), or (\ref{eq1.2''}) in terms of the 
entropy $S$, for both directions of the string 
$\rightleftharpoons$ black hole transition. In three spatial 
dimensions, we find that the size (computed in real
spacetime) of a {\it typical self-gravitating} string is given by 
the random walk value (\ref{eq1.5}) when $g^2 \le g_0^2$, with 
$g_0^2 \sim (M/M_s)^{-3/2} \sim S^{-3/2}$, and by
\begin{equation}
R_{\rm typ} \sim \frac{1}{g^2 \, M} \, , \label{eq1.8}
\end{equation}
when $g_0^2 \le g^2 \le g_c^2$. Note that $R_{\rm typ}$ smoothly
interpolates between $R_s^{\rm rw}$ and $\ell_s$. This result 
confirms the picture proposed by Ref.~\cite{hp2} when $d=3$, but 
with the bonus that Eq.~(\ref{eq1.8}) (which agrees with 
Eq.~(\ref{eq1.6bis})) refers to a radius which is estimated directly 
in physical space (see below), and which is the size of a typical 
member of the microcanonical ensemble of self-gravitating strings. 
In all higher dimensions\footnote{With the proviso that the 
consistency of our analysis is open to doubt when $d\geq 8$.}, we 
find that the size of a typical self-gravitating string remains 
fixed at the random walk value (\ref{eq1.5}) when $g \le
g_c$. However,  when $g$ gets close to a value of order $g_c$, the 
ensemble of self-gravitating strings becomes (smoothly in $d=4$, but 
suddenly in $d \geq 5$) dominated by very compact strings of size 
$\sim \ell_s$ (which are then expected to collapse with a slight 
further increase of $g$ because the dominant size is only slightly 
larger than the Schwarzschild radius at $g_c$).

Our results confirm and clarify the main idea of a correspondence
between string states and black hole states \cite{susskind},
\cite{halyo}, \cite{hp1}, \cite{hp2}, and suggest that the 
transition between these states is rather smooth, with no apparent 
hysteresis, and with continuity in entropy, mass and typical size. 
It is, however, beyond the technical grasp of our analysis to 
compute any precise number at the transition (such as the famous 
factor $1/4$ in the Bekenstein-Hawking entropy formula).

\section{Size distribution of free string states}
The aim of this section is to estimate the distribution function in
size of the ensemble of free string states of mass $M$, i.e. to 
count how many massive string states have a given size $R$. This 
estimate will be done while neglecting the gravitational 
self-interaction. The effect of the latter will be taken into 
account in a later section.

Let us first estimate the distribution in size by a rough, heuristic
argument based on the random walk model \cite{rw} of a generic 
excited string state. In string units ($\ell_s \sim M_s^{-1} \sim 
1$), the geometrical shape (in $d$-dimensional space) of a generic 
massive string state can be roughly identified with a random walk of 
$M$ steps of unit length. We can constrain this random walk to stay 
of size $\lsim R$ by considering a diffusion process, starting from 
a point source at the origin, in presence of an {\it absorbing} 
sphere $S_R$ of radius $R$, centered on the origin. In the 
continuous approximation, the kernel $K_t (\mbox{\boldmath$x$} , 
\mbox{\boldmath$0$})$ giving the conditional probability density of 
ending, at time $t$, at position $\mbox{\boldmath$x$}$, after having 
started (at time $0$) at the origin, without having ever gone 
farther from the origin than the distance $R$, satisfies: (i) the 
diffusion equation $\partial_t \, K_t = \Delta \, K_t$, (ii) the 
initial condition $K_0 (\mbox{\boldmath$x$} , \mbox{\boldmath$0$}) = 
\delta (\mbox{\boldmath$x$})$, and (iii) the ``absorbing'' boundary 
condition $K_t = 0$ on the sphere $S_R$. The kernel $K_t$ can be 
decomposed in eigenmodes,
\begin{equation}
K_t (\mbox{\boldmath$x$} , \mbox{\boldmath$0$}) = \sum_n \psi_n
(\mbox{\boldmath$x$}) \, \psi_n (\mbox{\boldmath$0$}) \, e^{-E_n \, 
t}\, , \label{eq2.1}
\end{equation}
where $\psi_n (\mbox{\boldmath$x$})$ is a normalized (real) $L^2$
basis ($\int d^d \, x \, \psi_n (\mbox{\boldmath$x$}) \, \psi_m
(\mbox{\boldmath$x$}) = \delta_{nm}$; $\displaystyle{\sum_{n}} \
\psi_n (\mbox{\boldmath$x$}) \, \psi_n (\mbox{\boldmath$y$}) = 
\delta (\mbox{\boldmath$x$} - \mbox{\boldmath$y$})$) satisfying
\begin{equation}
\Delta \, \psi_n = -E_n \, \psi \, , \label{eq2.2}
\end{equation}
in the interior, and vanishing on $S_R$. The total conditional
probability of having stayed within $S_R$ after the time $t$ is the
integral $\int_{B_R} d^d \, x \, K_t (\mbox{\boldmath$x$} ,
\mbox{\boldmath$0$})$ within the ball $B_R : \vert 
\mbox{\boldmath$x$}\vert \leq R$. For large values of $t$, $K_t$ is 
dominated by the lowest eigenvalue $E_0$, and the conditional 
probability goes like $c_0 \, e^{-E_0 \, t}$, where $c_0$ is a 
numerical constant of order unity. The eigenvalue problem 
(\ref{eq2.2}) is easily solved (in any dimension $d$), and the 
$s$-wave ground state can be expressed in terms of a
Bessel function: $\psi_0 (r) = N \, J_{\nu} (k_0 \, r) / (k_0 \,
r)^{\nu}$ with $\nu = (d-2)/2$. Here, $k_0 = \sqrt{E_0}$ is given by
the first zero $j_{\nu}$ of $J_{\nu} (z) : k_0 = j_{\nu} / R$. The
important information for us is that the ground state energy $E_0$
scales with $R$ like $E_0 = c_1 / R^2$, where $c_1 = {\cal O} (1)$ 
is a numerical constant. This scaling is evident for the Dirichlet 
problem(\ref{eq2.2}), whatever be the shape of the boundary. 
Finally, remembering that the number of time steps is given by the 
mass, $t=M$, we expect the looked for conditional probability, i.e. 
the {\it fraction} of all string states at mass level $M$ which are 
of size $\lsim R$, to be asymptotically of order
\begin{equation}
f(R) \sim e^{-c_1 \, M / R^2} \, . \label{eq2.3}
\end{equation}
This estimate is expected to be valid when $M / R^2 \gg 1$, i.e. for
string states which are much smaller (in size) than a typical random
walk $R_{\rm rw}^2 \sim M$ (but still larger than the string length,
$R \gsim 1$). In the opposite limit, $R^2 \gg M$, the kernel $K_t
(\mbox{\boldmath$x$}, \mbox{\boldmath$0$})$ can be approximated by 
the free-space value $K_t^{(0)} (\mbox{\boldmath$x$}, 
\mbox{\boldmath$0$})= (4 \, \pi \, t)^{-d/2} \, \exp 
(-\mbox{\boldmath$x$}^2 / (4t))$, with $t=M$, so that the fraction 
of string states of size $\gsim R \gg R_{\rm rw}$ will be of order 
$\sim e^{-c_2 \, R^2 / M}$, with $c_2 = {\cal O} (1)$.

As the result (\ref{eq2.3}) will be central to the considerations of
this paper, we shall now go beyond the previous heuristic, random 
walk argument and derive the fraction of small string states by a 
direct counting of quantum string states. For simplicity, we shall 
deal with open bosonic strings ($\ell_s \equiv \sqrt{2 \, \alpha'}$, 
$0 \leq \sigma \leq \pi$)
\begin{equation}
X^{\mu} (\tau , \sigma) = X_{\rm cm}^{\mu} (\tau , \sigma) +
\widetilde{X}^{\mu} (\tau , \sigma) \, , \label{eq2.4}
\end{equation}
\begin{equation}
X_{\rm cm}^{\mu} (\tau , \sigma) = x^{\mu} + 2 \, \alpha' \, p^{\mu} \,
\tau \, , \label{eq2.5}
\end{equation}
\begin{equation}
\widetilde{X}^{\mu} (\tau , \sigma) = i \, \ell_s \sum_{n \not= 0} \
\frac{\alpha_n^{\mu}}{n} \ e^{-i n \tau} \, \cos \, n \, \sigma \, .
\label{eq2.6}
\end{equation}
Here, we have explicitly separated the center of mass motion $X_{\rm
cm}^{\mu}$ (with $[x^{\mu} , p^{\nu}] = i \, \eta^{\mu \nu}$) from 
the oscillatory one $\widetilde{X}^{\mu}$ ($[\alpha_m^{\mu} ,
\alpha_n^{\nu}] = m \, \delta_{m+n}^0 \, \eta^{\mu \nu}$). The free
spectrum is given by $\alpha' \, M^2 = N-1$ where $(\alpha \cdot 
\beta\equiv \eta_{\mu \nu} \, \alpha^{\mu} \, \beta^{\nu} \equiv - 
\alpha^0\, \beta^0 + \alpha^i \, \beta^i)$
\begin{equation}
N = \sum_{n=1}^{\infty} \ \alpha_{-n} \cdot \alpha_n =
\sum_{n=1}^{\infty} \ n \, N_n \, . \label{eq2.7}
\end{equation}
Here $N_n \equiv a_n^{\dagger} \cdot a_n$ is the occupation number 
of the $n^{\rm th}$ oscillator ($\alpha_n^{\mu} = \sqrt{n} \ 
a_n^{\mu}$,$[a_n^{\mu} , a_m^{\nu \dagger}] = \eta^{\mu \nu} \, 
\delta_{nm}$, with $n,m$ positive).

The decomposition (\ref{eq2.4})--(\ref{eq2.6}) holds in any 
conformal gauge ($(\partial_{\tau} \, X^{\mu} \pm \partial_{\sigma} 
\, X^{\mu})^2 = 0$). One can further specify the choice of 
worldsheet coordinates by imposing
\begin{equation}
n_{\mu} \, X^{\mu} (\tau , \sigma) = 2 \alpha' (n_{\mu} \, p^{\mu})
\, \tau \, , \label{eq2.8}
\end{equation}
where $n^{\mu}$ is an arbitrary timelike or null vector ($n \cdot n
\leq 0$) \cite{scherk}. Eq.~(\ref{eq2.8}) means that the 
$n$-projected oscillators $n_{\mu} \, \alpha_m^{\mu}$ are set equal 
to zero. The usual ``light-cone'' gauge is obtained by choosing a 
fixed, null vector $n_{\mu}$. The light-cone gauge introduces a 
preferred (``longitudinal'') direction in space, which is an 
inconvenience for defining the (rms) size of a massive string state. 
As we shall be interested in quasi-classical, very massive string 
states ($N \gg 1$) it should be possible to work in the ``center of 
mass'' gauge, where the vector $n^{\mu}$ used in Eq.~(\ref{eq2.8}) 
to define the $\tau$-slices of the world-sheet is taken to be the 
total momentum $p^{\mu}$ of the string. This gauge is the most 
intrinsic way to describe a string in the classical limit. Using 
this intrinsic gauge, one can covariantly\footnote{In an arbitrary 
conformal gauge, the definition (\ref{eq2.9}) is gauge-dependent (in 
spite of the use of the orthogonal projection) because both the 
definition of $X_{\rm cm}^{\mu} (\tau)$, and that of the $(\sigma , 
\tau)$-averaging depend on the choice of world-sheet gauge. Even if 
we were using the (more intrinsic but more complicated) average with 
weight $\sqrt{-\det \, \gamma_{ab}}\ d \sigma \, d \tau = 
(\partial_{\sigma} \, X^{\mu})^2 \, d \sigma \, d\tau$, the 
dependence upon $X_{\rm cm}^{\mu} (\tau)$ would remain.}
define the proper rms size of a massive string state as
\begin{equation}
R^2 \equiv \frac{1}{d} \ \langle (\widetilde{X}_{\perp}^{\mu} \, 
(\tau, \sigma))^2 \rangle_{\sigma , \tau} \, , \label{eq2.9}
\end{equation}
where $\widetilde{X}_{\perp}^{\mu} \equiv \widetilde{X}^{\mu} -
p^{\mu} (p \cdot \widetilde{X}) / (p \cdot p)$ denotes the 
projection of $\widetilde{X}^{\mu} \equiv X^{\mu} - X_{\rm cm}^{\mu} 
(\tau)$ orthogonally to $p^{\mu}$, and where the angular brackets 
denote the (simple) average with respect to $\sigma$ and $\tau$. 
The factor $1/d$ in Eq.~(\ref{eq2.9}) is introduced to simplify 
later formulas. So defined $R$ is the rms value of the {\it projected 
size} of the string along an arbitrary, but fixed spatial direction. 
[We shall find that this projected size is always larger than 
$\sqrt{3 \, \alpha' / 2}$; i.e. string states cannot be 
``squeezed'', along any axis, more than this.]

In the center of mass gauge, $p_{\mu} \, \widetilde{X}^{\mu}$ 
vanishes by definition, and Eq.~(\ref{eq2.9}) yields simply
\begin{equation}
R^2 = \frac{1}{d} \, \ell_s^2 \, {\cal R} \, , \label{eq2.10}
\end{equation}
with
\begin{equation}
{\cal R} \equiv \sum_{n=1}^{\infty} \ \frac{\alpha_{-n} \cdot 
\alpha_n+ \alpha_n \cdot \alpha_{-n}}{2 \ n^2} \, . \label{eq2.11}
\end{equation}
The squared-size operator ${\cal R}$, Eq.~(\ref{eq2.11}), contains 
the logarithmically infinite contribution $\sum\, 1/(2n)$. Without 
arguing with the suggestion that this contribution may have a 
physical meaning (see, e.g., \cite{kks88}), we note here that this 
contribution is state-independent. We are interested in this 
work in the relative sizes of various highly excited, 
quasi-classical states. A concept which should reduce to the 
well-defined, finite rms size of a classical Nambu string in the 
classical limit. We shall therefore discard this state-independent 
contribution, i.e. work with the normal-ordered operator
\begin{equation}
:{\cal R}: \ = \sum_{n=1}^{\infty} \ \frac{a_n^{\dagger} \cdot 
a_n}{n}= \sum_{n=1}^{\infty} \ \frac{N_n}{n} \, . \label{eq2.12}
\end{equation}

We shall assume that we can work both in the center-of-mass
(worldsheet) gauge ($p_{\mu} \, \alpha_m^{\mu} \rightarrow 0$) and 
in the center-of-mass (Lorentz) frame ($(p^{\mu}) = (M ,
\mbox{\boldmath$0$})$). This means that the scalar product in the
level occupation number $N_n$ runs over the $d$ spatial dimensions: 
$N_n = a_n^{\dagger} \cdot a_n = \displaystyle{\sum_{i=1}^d} \
(a_n^i)^{\dagger} \, a_n^i$. The ``wrong sign'' time oscillators
$\alpha_n^0$ are set equal to zero. The Virasoro constraints then
imply, besides the mass formula $\alpha' \, M^2 = N-1$, the usual
sequence of constraints on physical states, $L_n \, \vert \phi 
\rangle= 0$, with $L_n = \frac{1}{2} \ \displaystyle{\sum_{m}^{}} \
\displaystyle{\sum_{i=1}^d} \ \alpha_{n-m}^i \, \alpha_m^i$. These
constraints mean that the $d$ oscillators $\alpha_n^i$ at level $n$
are not physically independent.

The problem we would like to solve is to count the number of 
physical states, in the Fock space of the center-of-mass oscillators
$\alpha_n^i$, having some fixed values of $N$ and ${\cal R}$ (we
henceforth work only with the normal-ordered operator (\ref{eq2.12})
without adorning it with the $: \ :$ notation). The Virasoro
constraints make this problem technically quite difficult. However, 
we know from the exact counting of physical states (without size
restriction) in the light-cone gauge that the essential physical
effect of the Virasoro constraints is simply to reduce the number of
independent oscillators at any level $n$ from $d = D-1$ (in the
center-of-mass gauge) to $d-1 = D-2$. If we (formally) consider $d$ 
as a large parameter\footnote{It would be interesting to see if one 
can technically implement a large $d$ approach to our counting 
problem.}, this change in the number of effective free oscillators 
should have only a small fractional effect on any other 
coarse-grained, counting problem. We shall assume that this is the 
case, and solve the much simpler counting problem where the $d$ 
oscillators $\alpha_n^i$ are considered as independent\footnote{We 
tried to work in the light-cone gauge, with $d-1$ independent 
oscillators. However, the necessary inclusion of the longitudinal 
term $M^{-2} ( p \cdot \widetilde{X})^2$in (\ref{eq2.9}), which is 
quadratic in the longitudinal oscillators $\alpha_n^- = (p^+)^{-1} 
\, L_n^{\rm transverse}$, leads to a complicated, interacting theory 
of the $d-1$ transverse oscillators.}. To solve this problem we pass 
from a microcanonical problem (fixed values of $N$ and ${\cal R}$) 
to a grand canonical one (fixed values of some thermodynamical 
conjugates of $N$ and ${\cal R}$). Let us introduce the formal 
``partition function''
\begin{equation}
Z_d (\beta , \gamma) \equiv \sum_{\{ N_n^i \}} \exp (-\beta \, N \,
[N_n^i] - \gamma \, {\cal R} \, [N_n^i]) \, , \label{eq2.13}
\end{equation}
where the sum runs over all sequences (labelled by $n \geq 1$ and $i 
=1, \ldots , d$) of independent occupation numbers $N_n^i =
(a_n^i)^{\dagger} \, a_n^i = 0,1,2,\ldots$, and where $N \, [N_n^i]$
and ${\cal R} \, [N_n^i]$ are defined by Eqs.~(\ref{eq2.7}), and
(\ref{eq2.12}), with $N_n \equiv \displaystyle{\sum_{i=1}^{d}} \,
N_n^i$. Note that (\ref{eq2.13}) is not the usual thermodynamical
partition function, and that $\beta$ is not the usual inverse
temperature. Indeed, $\beta$ is a formal conjugate to $N \simeq
\alpha' \, M^2$ and not to the energy $M$. In particular, because 
the degeneracy grows exponentially with $M$ (and not $M^2$) its 
Laplace transform (\ref{eq2.13}) is defined for arbitrary values of 
$\beta$. We associate with the definition (\ref{eq2.13}) that of a 
formal grand canonical ensemble of configurations, with the 
probability
\begin{equation}
p \, [\{ N_n^i \}] = Z_d^{-1} \, (\beta , \gamma) \, \exp (-\beta \, 
N\, [N_n^i] - \gamma \, {\cal R} \, [N_n^i]) \label{eq2.14}
\end{equation}
of realization of the particular sequence $N_n^i$ of occupation
numbers. The mean values of $N \, [N_n^i]$ and ${\cal R} \, [N_n^i]$
in this ensemble are
\begin{equation}
\overline{N} = - \frac{\partial \, \psi_d \, (\beta ,
\gamma)}{\partial \, \beta} \ , \ \overline{\cal R} = - 
\frac{\partial\, \psi_d \, (\beta , \gamma)}{\partial \, \gamma} \, 
, \label{eq2.15}
\end{equation}
where we denote
\begin{equation}
\psi_d \, (\beta , \gamma) \equiv \ln \, Z_d \, (\beta , \gamma) \, 
.\label{eq2.16}
\end{equation}
The second derivatives of the thermodynamical potential $\psi_d \,
(\beta , \gamma)$ give the fluctuations of $N$ and ${\cal R}$ in 
this grand canonical ensemble:
\begin{equation}
\overline{(\Delta N)^2} = \frac{\partial^2 \, \psi_d \, (\beta ,
\gamma)}{\partial \, \beta^2} \ , \ \overline{(\Delta {\cal R})^2} =
\frac{\partial^2 \, \psi_d \, (\beta , \gamma)}{\partial \, 
\gamma^2}\, . \label{eq2.17}
\end{equation}
Let us define as usual the entropy $S (\beta , \gamma)$ as the
logarithm of the number of string configurations having values of 
$N$and ${\cal R}$ equal to $\overline N$ and $\overline{\cal R}$,
Eqs.~(\ref{eq2.15}), within the precision of the rms fluctuations
(\ref{eq2.17}) \cite{landau}. This definition means that, in the
saddle-point approximation, $Z_d \, (\beta , \gamma) \simeq \exp \, 
[S- \beta \, \overline N - \gamma \, \overline{\cal R}]$, i.e.
\begin{equation}
\psi_d \, (\beta , \gamma) \simeq S (\beta , \gamma) - \beta \,
\overline N - \gamma \, \overline{\cal R} \, , \label{eq2.18}
\end{equation}
or
\begin{equation}
S \simeq \psi_d \, (\beta , \gamma) - \beta \, \frac{\partial \,
\psi_d \, (\beta , \gamma)}{\partial \, \beta} - \gamma \,
\frac{\partial \, \psi_d \, (\beta , \gamma)}{\partial \, \gamma} \, 
.\label{eq2.19}
\end{equation}
In other words, the entropy $S (\overline N , \overline{\cal R})$ is
the Legendre transform of $\psi_d \, (\beta , \gamma)$.

Because of the (assumed) independence of the $d$ oscillators in
(\ref{eq2.13}), one has
\begin{equation}
Z_d \, (\beta , \gamma) = \prod_{n=1}^{\infty} \ [1 - e^{-(\beta n +
\gamma / n)}]^{-d} \, , \label{eq2.19bis}
\end{equation}
i.e.
\begin{equation}
\psi_d \, (\beta , \gamma) = d \, \psi_1 \, (\beta , \gamma) \, ,
\label{eq2.20}
\end{equation}
with
\begin{equation}
\psi_1 \, (\beta , \gamma) = - \sum_{n=1}^{\infty} \ \ln \left[1 -
\exp \left( -\beta n - \frac{\gamma}{n} \right) \right] \, .
\label{eq2.21}
\end{equation}
We shall check {\it a posteriori} that we are interested in values 
of $\beta$ and $\gamma$ such that
\begin{equation}
\beta \ll \sqrt{\beta \, \gamma} \ll 1 \, . \label{eq2.22}
\end{equation}
For such values, one can approximate the discrete sum (\ref{eq2.21})
by a continuous integral over $x = \beta \, n$. This yields
\begin{equation}
\psi_1 \, (\beta , \gamma) = \frac{I (\delta)}{\beta} \ , \
\hbox{where} \ \delta \equiv \sqrt{\beta \, \gamma} \, ,
\label{eq2.23}
\end{equation}
and
\begin{equation}
I (\delta) \equiv - \int_0^{\infty} dx \, \ln \, [1 - e^{-(x \, + \,
\delta^2 / x)}] \, . \label{eq2.24}
\end{equation}
As $\delta = \sqrt{\beta \, \gamma} \ll 1$, we can try to expand
$I(\delta)$ in powers of $\delta$: $I(\delta) = I(0) + \delta \, 
I'(0)+ o (\delta)$. [Though the integral (\ref{eq2.24}) is expressed 
in terms of $\delta^2$, its formal expansion in powers of $\delta^2$
leads to divergent integrals.] The zeroth-order term is $I(0) = -
\int_0^{\infty} dx \, \ln \, (1-e^{-x}) = \pi^2 / 6$, while
$$
I'(0) = \lim_{\delta \rightarrow 0} \left[ -2 \int_0^{\infty}
\frac{du}{u} \ \frac{\delta}{e^{\delta (u+1/u)} - 1} \right] = -2
\int_0^{\infty} \frac{du}{u^2 + 1} = - \pi \, . \nonumber
$$
Hence, using (\ref{eq2.20}),
\begin{equation}
\psi_d \, (\beta , \gamma) = \frac{1}{\beta} \left[ C - D \,
\sqrt{\beta \, \gamma} + o \left( \sqrt{\beta \, \gamma} \right)
\right] \label{eq2.25}
\end{equation}
with
\begin{equation}
C = \frac{\pi^2}{6} \, d \ , \ D = \pi \, d \, . \label{eq2.26}
\end{equation}
[The notation $D$ in (\ref{eq2.26}) should not be confused with the
space-time dimension $d+1$.] The thermodynamic potential 
(\ref{eq2.25}) corresponds to the mean values
\begin{equation}
\overline N \simeq \frac{C - \frac{1}{2} \, D \, \sqrt{\beta \,
\gamma}}{\beta^2} \ , \ \overline{\cal R} \simeq \frac{D}{2 \, 
\delta}\, , \label{eq2.27}
\end{equation}
and to the entropy
\begin{equation}
S \simeq \frac{2 \, C - D \, \delta}{\beta} \simeq 2 \, \sqrt{C \,
\overline N} \left[ 1 - \frac{D^2}{8 \, C \, \overline{\cal R}}
\right] \, , \label{eq2.28}
\end{equation}
i.e.
\begin{equation}
S \simeq 2 \, \pi \, \sqrt{\frac{d}{6} \ \overline N} \left[ 1 -
\frac{3d}{4} \, \frac{1}{\overline{\cal R}} \right] \simeq 2 \, \pi
\left( \frac{d \, \alpha'}{6} \right)^{\frac{1}{2}} \, M \left[ 1 -
\frac{3}{4} \, \frac{\ell_s^2}{R^2} \right] \, . \label{eq2.29}
\end{equation}
The lowest-order term $2 \, \pi \, \sqrt{d \, \overline N / 6}$ is 
the usual (Hardy-Ramanujan) result for $d$ independent oscillators,
without size restriction. The factor in bracket, $1 - (3/4) 
(\ell_s^2/ R^2)$, with $\ell_s^2 = 2 \, \alpha'$, gives the 
fractional reduction in the entropy brought by imposing the size 
constraint $R^2\simeq d^{-1} \, \ell_s^2 \, \overline{\cal R}$. 
Under the conditions (\ref{eq2.22}) the fluctuations 
(\ref{eq2.17}) are fractionally small. More precisely, 
Eqs.~(\ref{eq2.17}) yield
\begin{equation}
\frac{\overline{(\Delta N)^2}}{\overline{N}^2} \sim \beta \sim 
\frac{M_s}{M} \ , \qquad \frac{\overline{(\Delta \, {\cal 
R})^2}}{\overline{\cal R}^2} \sim \frac{\beta}{\delta} \sim 
\frac{(R^2 / \ell_s^2)}{(M / M_s)} \, . \label{eq2.29bis}
\end{equation}

As said above, though we worked under the (physically incorrect) 
assumption of $d$ independent oscillators at each level $n$, we 
expect the result (\ref{eq2.28}) to be correct when $d \gg 1$. [We 
recall that the exact result for $S$ in absence of size restriction 
is $2 \, \pi \, \sqrt{(d-1) \, \overline N / 6}$.] Note the rough 
physical meanings of the auxiliary quantities $\beta$, $\gamma$ and 
$\delta$: $\beta \sim (\overline{N})^{-1/2} \sim (M/M_s)^{-1}$, 
$\delta \sim \overline{\cal R}^{-1} \sim \ell_s^2 / R^2$, $\gamma 
\sim (\overline{N})^{1/2} / \overline{\cal R}^2 \sim M\, \ell_s^5 / 
R^4$.

Summarizing, the main result of the present section is that the 
number(``degeneracy'') of free string states of mass $M$ and size 
$R$ (within the narrow bands defined by the fluctuations 
(\ref{eq2.17})) is of the form
\begin{equation}
{\cal D} \, (M,R) \sim \exp \, [ c \, (R) \, a_0 \, M ] \, ,
\label{eq2.30}
\end{equation}
where $a_0 = 2 \, \pi \, ((d-1) \, \alpha' / 6)^{1/2}$ and
\begin{equation}
c \, (R) = \left( 1 - \frac{c_1}{R^2} \right) \left( 1 - c_2 \,
\frac{R^2}{M^2} \right) \, , \label{eq2.31}
\end{equation}
with the coefficients $c_1$ and $c_2$ being of order unity in string
units. [We have added, for completeness, in Eq.~(\ref{eq2.30}) the
factor $1 - c_2 \, R^2 / M^2$ which operates when one considers very
``large'' string states, $R^2 \gg R_{\rm rw}^2$ (as discussed 
below Eq.~(\ref{eq2.3})).] The coefficient $c \, (R)$ gives the 
fractional reduction in entropy brought by imposing a size 
constraint. Note that (as expected) this reduction is minimized when 
$c_1 \, R^{-2} \sim c_2 \, R^2 / M^2$, i.e. for $R \sim R_{\rm rw} 
\sim \ell_s \, \sqrt{M/M_s}$. [The absolute reduction in degeneracy 
is only a factor ${\cal O} (1)$ when $R \sim R_{\rm rw}$.] Note also 
that $c \, (R) \rightarrow 0$ both when $R \sim \ell_s$ and when $R 
\sim \ell_s (M/M_s)$. [The latter value corresponding to the 
vicinity of the leading Regge trajectory $J\sim \alpha' \, M^2$.]

\section{Mass shift of string states due to self-gravity}

In this section we shall estimate the mass shift of string states 
(of mass $M$ and size $R$) due to the exchange of the various 
long-range fields which are universally coupled to the string: 
graviton, dilaton and axion. As we are interested in very massive 
string states, $M \gg M_s$, in extended configurations, $R \gg 
\ell_s$, we expect that massless exchange dominates the 
(state-dependent contribution to the) mass shift.

The evaluation, in string theory, of (one loop) mass shifts for
massive states is technically quite involved, and can only be 
tackled for the states which are near the leading Regge trajectory
\cite{mshift}. [Indeed, the vertex operators creating these states 
are the only ones to admit a manageable explicit oscillator
representation.] As we consider states which are very far from the
leading Regge trajectory, there is no hope of computing exactly (at
one loop) their mass shifts. We shall resort to a semi-classical
approximation, which seems appropriate because we consider highly
excited configurations. As a starting point to derive the mass-shift
in this semi-classical approximation we shall use the classical
results of Ref.~\cite{BD98} which derived the effective action of
fundamental strings. The one-loop exchange of $g_{\mu \nu}$, 
$\varphi$ and $B_{\mu \nu}$ leads to the effective action
\begin{equation}
I^{\rm eff} = I_0 + I_1 \, , \label{eq3.1}
\end{equation}
where $I_0$ is the free (Nambu) string action ($d^2 \, \sigma_1 
\equiv d \sigma_1 \, d \tau_1$, $\gamma_1 = -\det \, \gamma_{ab} \, 
(X^{\mu}(\sigma_1 , \tau_1))$)
\begin{equation}
I_0 = -T \int d^2 \, \sigma_1 \, \sqrt{\gamma_1} \, , \label{eq3.2}
\end{equation}
and $I_1$ the effect of the one-loop interaction $(X_1^{\mu} \equiv
X^{\mu} (\sigma_1 , \tau_1), \ldots)$
\begin{equation}
I_1 = 2\pi \int \int d^2 \sigma_1 \, d^2 \sigma_2 \, G_F (X_1 
-X_2) \, \sqrt{\gamma_1} \, \sqrt{\gamma_2} \ C_{\rm tot} (X_1 , 
X_2) \,, \label{eq3.3}
\end{equation}
where $G_F$ is Feynman's scalar propagator ($\Box \, G_F (x) =
-\delta^D (x)$), and $C_{\rm tot} (X_1 , X_2) = J_{\varphi} (X_1) 
\cdot J_{\varphi} (X_2) + J_g (X_1) \cdot J_g (X_2) + J_B (X_1) 
\cdot J_B (X_2)$ comes from the couplings of $\varphi$, $g_{\mu 
\nu}$ and $B_{\mu \nu}$ to their corresponding world-sheet sources 
(indices suppressed; spin-structure hidden in the dot product). The 
exchange term $C_{\rm tot}$ takes, in null (conformal) coordinates 
$\sigma^{\pm} = \tau \pm\sigma$, the simple left-right factorized 
form \cite{BD98}
\begin{equation}
\sqrt{\gamma_1} \, \sqrt{\gamma_2} \ C_{\rm tot} (X_1 , X_2) = 32 \, 
G_N\, T^2 \, (\partial_+ \, X_1^{\mu} \ \partial_+ \, X_{2 \mu})
(\partial_- \, X_1^{\nu} \ \partial_- \, X_{2 \nu}) \, . 
\label{eq3.4}
\end{equation}
Here, $T = (2 \, \pi \, \alpha')^{-1}$ is the string tension, 
$G_N$ is Newton's constant\footnote{Normalized, in any dimension, by 
writing the Einstein action as $(16 \, \pi \, G_N)^{-1} \int d^D \, 
x \, \sqrt{g} \, R(g)$.} and $\partial_{\pm} = \partial / \partial 
\sigma^{\pm} = \frac{1}{2} \, (\partial_{\tau} \pm 
\partial_{\sigma})$. Let us define $P_{\pm}^{\mu} = P_{\pm}^{\mu}
(\sigma^{\pm})$ by ($\ell_s = \sqrt{2 \, \alpha'}$ as above)
\begin{equation}
2 \, \partial_{\pm} \, X^{\mu} = \ell_s \, P_{\pm}^{\mu} \, ,
\label{eq3.5}
\end{equation}
so that,  for an open (bosonic) string (with $\alpha_0^{\mu} \equiv
\ell_s \, p^{\mu}$),
\begin{equation}
P_{\pm}^{\mu} = \sum_{-\infty}^{+\infty} \ \alpha_n^{\mu} \, e^{-i n
\sigma^{\pm}} \, . \label{eq3.6}
\end{equation}
Using the definition (\ref{eq3.5}) and inserting the Fourier
decomposition of $G_F$ yields
\begin{eqnarray}
I_1 &= &\frac{4 \, G_N}{\pi} \int \frac{d^D \, k}{(2\pi)^D} \,
\frac{1}{k^2 - i \, \varepsilon} \int \int d^2 \sigma_1 \, d^2 
\sigma_2\, (P_+ (X_1) \cdot P_+ (X_2)) (P_- (X_1) \cdot P_- (X_2)) 
\nonumber \\
&\times &e^{ik \cdot (X_1 - X_2)} \, , 
\label{eq3.7}
\end{eqnarray}
where one recognizes the insertion of two gravitational vertex 
operators $V^{\mu \nu} (k;X) = P_+^{\mu} (X) \, P_-^{\nu} (X) \, 
e^{ik \cdot X}$ at two different locations on the worldsheet, and 
with two opposite momenta for the exchanged graviton\footnote{For 
simplicity, we call ``graviton'' the exchanged particle, which is a 
superposition of the graviton, the dilaton and the axion.}. [Note 
that the exchanged graviton is off-shell.] It is convenient to use 
the Virasoro constraints $(P_{\pm}^{\mu} (X))^2 = 0$ to replace in 
Eq.~(\ref{eq3.6}) $P_{\pm} (X_1) \cdot P_{\pm} (X_2) = - \frac{1}{2} 
\, (\Delta \, P_{\pm}^{\mu})^2$where $\Delta \, P_{\pm}^{\mu} \equiv 
P_{\pm}^{\mu} (X_1) - P_{\pm}^{\mu} (X_2)$. It is important to note 
that the zero mode contribution $\alpha_0^{\mu}$ drops out of 
$\Delta \, P_{\pm}^{\mu}$ (i.e. $\Delta \, P_{\pm}^{\mu} = \Delta \, 
\widetilde{P}_{\pm}^{\mu}$ is purely oscillatory).

Writing that the correction $I_1$ to the effective action $I^{\rm 
eff}$ (which gives the vacuum persistence amplitude; see, e.g., 
Eq.~(7) of \cite{BD98}) must correspond to a phase shift $- \int 
\delta E \, dt = - \int \delta M \, d X_{\rm cm}^0$, in the 
center-of-mass frame of the string, yields (with the normalization 
(\ref{eq2.5})) the link $I_1 = -\ell_s^2 \int d \tau \, M \, \delta M 
= - \frac{1}{2} \, \ell_s^2 \int d \tau \, \delta M^2$. Let us also 
define $\Delta \, X^{\mu} \equiv X_1^{\mu} - X_2^{\mu}$ and 
decompose it in its zero-mode part $\Delta \, X_{\rm cm}^{\mu} = 
\ell_s^2 \, p^{\mu} (\tau_1 - \tau_2)$ and its oscillatory part 
$\Delta \, \widetilde{X}^{\mu} = \widetilde{X}_1^{\mu} - 
\widetilde{X}_2^{\mu}$. Finally, the mass-shift can be read from
\begin{equation}
\int d\tau \, \delta M^2 = - \frac{2 \, G_N}{\pi \, \ell_s^2} \int
\frac{d^D \, k}{(2\pi)^D} \, \frac{1}{k^2 - i \varepsilon} \int \int 
d^2 \sigma_1 \, d^2 \sigma_2 \, e^{i \, \ell_s^2 \, k \cdot p \, 
(\tau_1 -\tau_2)} \, W (k,1,2) \, , \label{eq3.8}
\end{equation}
where ($1$ and $2$ being short-hands for $(\tau_1 , \sigma_1)$ and
$(\tau_2 , \sigma_2)$, respectively)
\begin{equation}
W (k,1,2) = (\Delta \, P_+^{\mu} (1,2))^2 \, (\Delta \, P_-^{\nu}
(1,2))^2 \, e^{ik \cdot \Delta \widetilde{X} (1,2)} \, . 
\label{eq3.9}
\end{equation}
Interpreted at the quantum level, the classical result (\ref{eq3.8})
gives (modulo some ordering problems, which are, however, 
fractionally negligible when considering very massive states) the 
mass-shift $\delta\, M_N^2$ of the string state $\vert N \rangle$ 
when replacing $W (k,1,2)$, on the right-hand side of 
Eq.~(\ref{eq3.8}), by the quantum average $\langle N \vert \, W 
(k,1,2) \, \vert N \rangle$. Here, we shall mainly be interested in 
the real part of $\delta \, M^2$, obtained by replacing $(k^2 - i 
\varepsilon)^{-1}$ by the principal part of $(k^2)^{-1}$ (denoted 
simply $1/k^2$), i.e. the Feynman Green's function $G_F (x)$ by the 
half-retarded-half-advanced one $G_{\rm sym} (x)$. [The imaginary 
part of $\delta \, M^2$ gives the decay rate, i.e. the total rate of 
emission of massless quanta.] As $L_0 - 1$ is the ``Hamiltonian'' 
that governs the $\tau$-evolution of an open string, 
the vanishing of $(L_0 - 1) \, \vert N \rangle$ for any physical 
state ensures that $\langle N \vert \, W (k,1,2) \, \vert N \rangle$ 
is $\tau$-translation invariant, i.e. that it depends only on the
difference $\tau_{12} \equiv \tau_1 - \tau_2$, and not on the 
average $\overline{\tau} \equiv \frac{1}{2} \, (\tau_1 + \tau_2)$. 
This means that the double world-sheet integration $d^2 \sigma_1 \, 
d^2 \sigma_2 = d \tau_1 \, d \sigma_1 \, d \tau_2 \, d \sigma_2 = d 
\overline{\tau} \, d\tau_{12} \, d \sigma_1 \, d \sigma_2$ 
on the right-hand side of Eq.~(\ref{eq3.8}) contains a formally 
infinite infra-red ``volume'' factor $\int d \, \overline{\tau}$ 
which precisely cancels the integral $\int d \, \tau$ on the 
left-hand side to leave a finite answer for $\delta \, M^2$.

It is also important to note the good ultraviolet behaviour of
Eq.~(\ref{eq3.8}). The ultraviolet limit $k \rightarrow \infty$
corresponds to the coincidence limit $(\tau_2 , \sigma_2) 
\rightarrow(\tau_1 , \sigma_1)$ on the world-sheet. Let us define $u 
\equiv \sigma_1^+ - \sigma_2^+$, $v \equiv \sigma_1^- - \sigma_2^-$ 
and consider the coincidence limit $u \rightarrow 0$, $v \rightarrow 
0$. In this limit the vertex insertion factors $(\Delta \, P_+)^2 \, 
(\Delta \, P_-)^2$ tend to zero like $u^2 \, v^2$, while the Green's 
function blows up like $[(\Delta \, X)^2]^{-(D-2)/2}
 \propto (uv)^{-(D-2)/2}$. The resulting integral, $\int du \, dv \, 
(uv)^{-(D-6)/2}$, has its first ultraviolet pole when the space-time 
dimension $D = d+1 = 8$. This means probably that in dimensions $D 
\geq 8$ the exchange of massive modes (of closed strings) becomes 
important. Our discussion, which is limited to considering only the 
exchange of massless modes, is probably justified only when $D < 8$.

Following the (approximate) approach of Section~2 we shall estimate 
the average mass shift $\delta \, M^2 (R)$ for string states of size 
$R$ by using the grand canonical ensemble with density matrix
\begin{equation}
\rho \equiv (Z_d \, (\beta , \gamma))^{-1} \, \exp \, (-\beta \, N 
\, -\gamma \, : {\cal R} : \, ) \, , \label{eq3.10}
\end{equation}
where the operators $N$ and $: {\cal R} :$, defined by
Eqs.~(\ref{eq2.7}) and (\ref{eq2.12}), belong to the Fock space 
built upon $d$ sequences of string oscillators $\alpha_n^i$ (formal
``center-of-mass'' oscillators). For any quantity $Q$ (built from 
string oscillators) we denote the grand canonical average as 
$\langle Q \rangle_{\beta , \gamma} \equiv {\rm tr} \, (Q \, \rho)$. 
Using the $\tau$-shift invariance mentioned above, Eq.~(\ref{eq3.8}) 
yields
\begin{equation}
\delta \, M_{\beta , \gamma}^2 = - \frac{2 \, G_N}{\pi \, \ell_s^2} 
\int \frac{d^d \, k \ d \, \omega}{(2\pi)^{d+1} \, 
[\mbox{\boldmath$k$}^2 -\omega^2 - i \varepsilon]} \int d \tau_{12} 
\, d \sigma_1 \, d \sigma_2\, e^{-i \, \ell_s^2 M \omega \, 
\tau_{12}} \, \langle W (k,1,2) \rangle_{\beta , \gamma} \, , 
\label{eq3.11}
\end{equation}
where we have separated $k^{\mu}$ in its center-of-mass components 
$k^0= \omega$, $k^i = \mbox{\boldmath$k$}$, and where $\tau_{12} 
\equiv \tau_1 - \tau_2$ as above.

We shall estimate the grand canonical average $\langle W 
\rangle_{\beta, \gamma}$ in a semi-classical approximation in which 
we neglect some of the contributions linked to the ordering of the 
operator $W$, but take into account the quantum nature of the d
ensity matrix $\rho$, Eq.~(\ref{eq3.10}). The discreteness of the 
Fock states built from the$(a_n^i)^{\dagger}$, and the Planckian 
nature of $\rho$ will play a crucial role in the calculation below. 
[By contrast, a purely classical calculation would be awkward and 
ill-defined because of the problem of defining a measure on 
classical string configurations, and because of the Rayleigh-Jeans 
ultraviolet catastrophe.] To compute $\langle W \rangle_{\beta , 
\gamma}$ it is convenient to define it as a double contraction of 
the coefficient of $\zeta_{\mu_1}^1$ $\zeta_{\mu_2}^2$ 
$\zeta_{\mu_3}^3$ $\zeta_{\mu_4}^4$ in the exponentiated version of 
$W$:
\begin{equation}
W_{\zeta} \equiv : \exp \, [ \zeta_{\mu_1}^1 \, \Delta \, 
P_+^{\mu_1} +\zeta_{\mu_2}^2 \, \Delta \, P_+^{\mu_2} + 
\zeta_{\mu_3}^3 \, \Delta \,P_-^{\mu_3} + \zeta_{\mu_4}^4 \, \Delta 
\, P_-^{\mu_4} + ik \cdot \Delta \, \widetilde{X}] : \label{eq3.12}
\end{equation}
We shall define our ordering of $W$ by working with the normal 
ordered exponentiated operator (\ref{eq3.12}) (and picking the term 
linear in$\zeta_{\mu_1}^1$ $\zeta_{\mu_2}^2$ $\zeta_{\mu_3}^3$
$\zeta_{\mu_4}^4$). The average $\langle W_{\zeta} \rangle_{\beta} =
{\rm tr} \, (W_{\zeta} \, \rho)$ (where, to ease the notation, we 
drop the extra label $\gamma$) can be computed by a generalization 
of Bloch's theorem. Namely, if $A$ denotes any operator which is 
linear in the oscillators $\alpha_n^i$, we have the results
\begin{equation}
\langle e^A \rangle_{\beta} = \exp \left[ \frac{1}{2} \, \langle A^2
\rangle_{\beta} \right] \ ; \ \langle \, : e^A : \, \rangle_{\beta} 
=\exp \left[ \frac{1}{2} \, \langle \, : A^2 : \, \rangle_{\beta} 
\right] \, , \label{eq3.13}
\end{equation}
as well as their corollaries 
\begin{equation}
\langle e^A \rangle_0 = \exp \left[ \frac{1}{2} \, \langle A^2 
\rangle_0 \right] \ ; \ e^A = \, : e^A : \,\exp \left[ \frac{1}{2} 
\, \langle A^2 \rangle_0 \right] \, , \label{eq3.13bis}
\end{equation}
where $\langle W \rangle_0$ denotes the vacuum average (obtained in 
the zero temperature limit $\beta^{-1} \rightarrow 0$). The simplest 
way to prove these results is to use coherent-state methods 
\cite{AABO} (see also,Ref.~\cite{scherk} and the Appendix~7.A of 
Ref.~\cite{GSW}). For instance, to prove the second equation 
(\ref{eq3.13}) it is sufficient to consider a single oscillator and 
to check that (denoting $q = e^{-\epsilon}$, with $\epsilon = \beta 
\, n + \gamma / n$ (label $n$ suppressed), so that $Z = (1-q)^{-1}$, 
and $\vert b ) \equiv \exp \, (ba^{\dagger}) \, \vert 0 \rangle$)
\begin{eqnarray}
\langle e^{c_1 a^{\dagger}} \, e^{c_2 a} \rangle_{\beta} &=& Z^{-1} 
\,{\rm tr} \, (e^{c_1 a^{\dagger}} \, e^{c_2 a} \, q^{a^{\dagger} 
a}) =Z^{-1} \int \frac{d^2 \, b}{\pi} \, e^{-b^* b} (b \vert \, 
e^{c_1 a^{\dagger}} \, e^{c_2 a} \, q^{a^{\dagger} a} \, \vert b) 
\nonumber \\
&=& (1-q) \int \frac{d^2 \, b}{\pi} \, e^{-b^* b} (b \vert \, e^{c_1
b^*} \, e^{c_2 a} \, \vert qb) \nonumber \\
&=& (1-q) \int \frac{d^2 \, b}{\pi} \, e^{-(1-q) b^* b} \, e^{c_1 
b^* +c_2 qb} = e^{c_1 c_2 q / (1-q)} \, ,
\label{eq3.14}
\end{eqnarray}
and to recognize that $q / (1-q) = [e^{\epsilon} - 1]^{-1}$ is the
Planckian mean occupation number $\langle a^{\dagger} \, a
\rangle_{\beta}$.

If we apply the second Eq.~(\ref{eq3.13}) to an expression of the 
type$W_{\zeta} = \, : \exp \left( \displaystyle{\sum_{i=1}^{4}} \, 
\zeta_i\, A_i + B \right):$, one gets a Wick-type expansion for the 
coefficient (say $W_{1234}$) of $\zeta_1 \, \zeta_2 \, \zeta_3 \, 
\zeta_4$:
\begin{eqnarray}
W_{1234} &=& e^{\frac{1}{2} [BB]} \, ([A_1 A_2] \, [A_3 A_4] + 
\hbox{2 terms} + [A_1 B] \, [A_2 B] \, [A_3 A_4] + \hbox{5 terms} 
\nonumber \\
&+& [A_1 B] \, [A_2 B] \, [A_3 B] \, [A_4 B]) \, , \label{eq3.15}
\end{eqnarray}
where $[AB]$ denotes the ``thermal'' contraction $[AB] \equiv 
\langle \, : AB : \, \rangle_{\beta}$.

The looked-for grand canonical average of $W (k,1,2)$,
Eq.~(\ref{eq3.9}), is given by replacing $B = ik \cdot \Delta \,
\widetilde{X}$ and $A_1 = A_2 = \Delta \, P_+^{\mu}$, $A_3 = A_4 = 
\Delta \, P_-^{\nu}$ in Eq.~(\ref{eq3.15}). This leads to
\begin{equation}
\langle W \rangle_{\beta , \gamma} = e^{-\frac{1}{2} \langle : (k 
\cdot\Delta \widetilde{X})^2 : \rangle_{\beta , \gamma}} \left\{ 
\langle \, : (\Delta \, P_+)^2 : \, \rangle_{\beta , \gamma} \, 
\langle \, : (\Delta\, P_-)^2 : \, \rangle_{\beta , \gamma} + \ldots 
\right\} \, , \label{eq3.16}
\end{equation}
where the ellipsis stand for other contractions (which will be seen 
below to be subleading).

The calculation of the various contractions $[AB] \equiv \langle \, 
: AB : \, \rangle_{\beta}$ in Eqs.~(\ref{eq3.15}), (\ref{eq3.16}) is 
easily performed by using the basic contractions among the 
oscillators $a_n$,$a_m^{\dagger}$ ($n,m > 0$) (which are easily 
derived from the definition (\ref{eq3.10}) of the density matrix)
\begin{equation}
\langle \, : a_n^i \, (a_m^j)^{\dagger} : \, \rangle_{\beta , 
\gamma} =\langle \, : (a_m^j)^{\dagger} \, a_n^i : \, \rangle_{\beta 
, \gamma} =\frac{\delta^{ij} \, \delta_{nm}}{e^{\epsilon_n} - 1} \, 
, \label{eq3.17}
\end{equation}
where $\epsilon_n = \beta \, n + \gamma / n$. The other contractions
$[aa]$ and $[a^{\dagger} a^{\dagger}]$ vanish. In terms of the
$\alpha$-oscillators, the basic contraction reads $[\alpha_n^i \,
\alpha_m^j] = \delta^{ij} \, \delta_{n+m}^0 \, \vert n \vert / (\exp
(\epsilon_{\vert n \vert}) - 1)$, where now $n$ and $m$ can be 
negative(but not zero). Using these basic contractions, and the 
oscillator expansion (\ref{eq2.6}) of $\widetilde{X}^{\mu}$ (and 
noting that, in the center-of-mass frame only the spatial components 
of $\widetilde{X}^{\mu}$ survive) one gets
\begin{equation}
\langle \, : (\mbox{\boldmath$k$} \cdot \Delta \,
\widetilde{\mbox{\boldmath$X$}})^2 : \, \rangle_{\beta , \gamma} = 2 
\,\mbox{\boldmath$k$}^2 \, \ell_s^2 \ \sum_{n=1}^{\infty} \ 
\frac{x_n (1,2)}{n (e^{\epsilon_n} - 1)} \, , \label{eq3.18}
\end{equation}
with
\begin{equation}
x_n \, (1,2) = \cos^2 \, n \, \sigma_1 + \cos^2 \, n \, \sigma_2 - 2 
\,\cos \, n \, \sigma_1 \, \cos \, n \, \sigma_2 \, \cos \, n \, 
\tau_{12} \, . \label{eq3.19}
\end{equation}
Similarly, the oscillator expansion (\ref{eq3.6}) yields
\begin{equation}
\langle \, : \Delta \, P_+)^2 : \, \rangle_{\beta , \gamma} = 4 \, d 
\ \sum_{n=1}^{\infty} \ \frac{n \, p_n^+ (1,2)}{e^{\epsilon_n} - 1} 
\, ,\label{eq3.20}
\end{equation}
with
\begin{equation}
p_n^+ (1,2) = 1 - \cos \, n (\sigma_1^+ - \sigma_2^+) = 1 - \cos \, 
n(\tau_{12} + \sigma_1 - \sigma_2) \, . \label{eq3.21}
\end{equation}
The result for $(\Delta \, P_-)^2$ is obtained by changing $\sigma^+
\rightarrow \sigma^-$ in Eq.~(\ref{eq3.21}) (i.e. $\sigma_1 - 
\sigma_2\rightarrow - \sigma_1 + \sigma_2$).

We can estimate the values of the right-hand sides of 
Eqs.~(\ref{eq3.18}) and (\ref{eq3.20}) by using the following 
``statistical'' approximation. In the parameter range discussed in 
Section~2 the basic sums $\sum \,n^{\pm 1} \, (e^{\epsilon_n} - 
1)^{-1}$, appearing in (\ref{eq3.18}), (\ref{eq3.20}), see their 
values dominated by a large interval, $\Delta\, n \gg 1$, around 
some $n_0 \gg 1$, of values of $n$, so that one can, with a good 
approximation, replace the discrete sum over $n$ by a formal
continuous integral over a real parameter. In such a continuous
approximation one can integrate by parts to show that any 
``oscillatory'' integral of the type $\int dn \, f(n) \, \cos \, n 
\, \sigma = [n^{-1} \, f(n) \, \sin \, n \, \sigma] - \int dn \, 
n^{-1} \, f' (n) \, \sin \, n\, \sigma$ is, because of the factors 
$n^{-1}$, numerically much smaller than the non-oscillatory one 
$\int dn \, f(n)$. [Here $\sigma$ denotes some combination of 
$\sigma_1$ and $\sigma_2$, like $2 \sigma_1$, $2 \sigma_2$, 
$\sigma_1 \pm \sigma_2$.] Alternatively, we can say that, 
for generic values of $\sigma_1$ and $\sigma_2$, one can treat in
Eqs.~(\ref{eq3.19}) or (\ref{eq3.21}) $\cos \, n \, \sigma_1$ and 
$\cos\, n \, \sigma_2$ as statistically independent random variables 
with zero average. Within such an approximation one can estimate 
(\ref{eq3.18}) by replacing $x_n (1,2)$ by $1$ (because $\cos^2 \, n 
\, \sigma_1 + \cos^2\, n \, \sigma_2 = 1 + \frac{1}{2} \, (\cos \, 2 
n \, \sigma_1 + \cos \, 2 n \, \sigma_2)$). Similarly, one can 
estimate (\ref{eq3.20}) by replacing $p_n^{\pm} \rightarrow 1$. The 
resulting estimates of (\ref{eq3.18}) and (\ref{eq3.20}) introduce 
exactly the grand canonical averages of the quantities $: {\cal R} 
:$ and $N$:
\begin{equation}
\frac{1}{2} \ \langle \, : (\mbox{\boldmath$k$} \cdot \Delta \,
\widetilde{\mbox{\boldmath$X$}})^2 : \, \rangle_{\beta , \gamma} 
\simeq\mbox{\boldmath$k$}^2 \, \langle R^2 \rangle_{\beta , \gamma} 
\, , \label{eq3.22}
\end{equation}
\begin{equation}
\langle \, : (\Delta \, P_+)^2 : \, \rangle_{\beta , \gamma} \simeq
\langle \, : (\Delta \, P_-)^2 : \, \rangle_{\beta , \gamma} \simeq 
4 \, \langle N \rangle_{\beta , \gamma} \simeq 4 \, \alpha' \, M^2 
\, . \label{eq3.23}
\end{equation}
Furthermore, one can check that the other contractions (like 
$[\Delta \, P_+ \, \Delta \, P_-]$ or $[\Delta \, P_+ \ k \cdot 
\Delta \, \widetilde{X}]$) entering Eq.~(\ref{eq3.16}) are all of 
the ``oscillatory'' type which is expected to give subleading 
contributions.

Inserting the results (\ref{eq3.22}), (\ref{eq3.23}) into
Eqs.~(\ref{eq3.16}) and (\ref{eq3.11}) leads to a trivial integral 
over $\tau_{12}$ ($\int d \, \tau_{12} \, \exp (-i \, \ell_s^2 \, M 
\, \omega \, \tau_{12}) = 2 \pi \, \delta (\omega) / (\ell_s^2 \, 
M)$) and, hence, to the following result for $\delta \, M = \delta 
\, M^2 / (2M)$
\begin{equation}
\delta \, M_{\beta , \gamma} \simeq - 4 \, \pi \, G_N \, M^2 \int
\frac{d^d \, k}{(2\pi)^d} \, \frac{e^{- \mbox{\boldmath$k$}^2
{\textstyle R^2}}}{\mbox{\boldmath$k$}^2 - i \, \varepsilon} \, . 
\label{eq3.24}
\end{equation}

The imaginary part of $\delta \, M$ is easily seen to vanish in the
present approximation. Finally,
\begin{equation}
\delta \, M_{\beta , \gamma} \simeq - c_d \, G_N \, 
\frac{M^2}{R^{d-2}}\, , \label{eq3.25}
\end{equation}
with the (positive\footnote{The sign $\delta \, M < 0$ was 
classically clear (even when taking into account relativistic 
effects), say in $d=3$, from the starting formulas (\ref{eq3.3}), 
(\ref{eq3.4}) where $G_{\rm sym} (x) = (4\pi)^{-1} \, \delta (x^2) > 
0$ and $4 (\partial_+ \, X_1 \cdot \partial_+ \, X_2) \, (\partial_- 
\, X_1 \cdot\partial_- \, X_2) = (\partial_+ \, \Delta \, X)^2 \, 
(\partial_- \, \Delta \, X)^2 > 0 $ because 
$\partial_{\pm}\Delta \, X^{\mu}$ is 
purely spacelike in the center-of-mass frame. The same conclusion 
would hold in the light-cone gauge.}) numerical constant
\begin{equation}
c_d = \left[ \frac{d-2}{2} \, (4\pi)^{\frac{d-2}{2}} \right]^{-1} \, 
,\label{eq3.26}
\end{equation}
equal to $1 / \sqrt{\pi}$ in $d=3$.

The result (\ref{eq3.25}) was expected in order of magnitude, but we
found useful to show how it approximately comes out of a detailed
calculation of the mass shift which incorporates both relativistic 
and quantum effects. It shows clearly that perturbation theory 
breaks down, even at arbitrarily small coupling, for sufficiently 
heavy and compact strings. Let us also point out that one can give a 
simple statistical interpretation of the calculation (\ref{eq3.15}) 
of the normal-ordered vertex operator $W (k,1,2)$, with the basic 
contractions (\ref{eq3.17}). The result of the calculation would 
have been the same if we had simply assumed that the oscillators 
$a_n^i$ were classical, complex random variables with a Gaussian 
probability distribution $\propto \exp \left[ - \frac{1}{2} \, 
(e^{\epsilon_n} - 1) \, \vert a_n^i \vert^2 \right]$. This 
equivalence underlies the success of the classical random walk model 
of a generic excited string state. The fact that the random walk 
must be made of $M/M_s$ independent steps is linked to the fact that 
the Planckian distribution of mean occupation numbers,
$\overline{N}_n = (\exp (\beta \, n + \gamma / n) -1)^{-1}$ is 
sharply cut off when $n \gsim \beta^{-1}$, i.e., from 
Eq.~(\ref{eq2.27}), when $n \gsim M/M_s$. More precisely, using the 
same ``statistical'' approximation as above, one finds that the slope
correlator $\langle \, : \partial_{\sigma} \, \widetilde{X}^i (\tau , 
\sigma_1) \, \partial_{\sigma} \, 
\widetilde{X}^j (\tau , \sigma_2) : \, \rangle_{\beta , \gamma}$ decays 
quite fast when $\vert \sigma_2 - \sigma_1 \vert \gsim 
M_s / M$.

Finally, let us mention that, by using the same tools as above, one 
can compute the imaginary part of the mass shift $\delta \, M = 
\delta \,M_{\rm real} - i \, \Gamma / 2$, i.e. the total decay rate 
$\Gamma$ in massless quanta. The quantity $\Gamma$ is, in fact, 
easier to define rigorously in string theory because, using $(k^2 - 
i \, \varepsilon)^{-1} = PP (k^2)^{-1} + i \pi \, \delta (k^2)$ in 
(\ref{eq3.8}), it is given by an integral where the massless quanta 
are all on-shell. When $\gamma = 0$ (a consistent approximation for 
a result dominated by $n \sim \beta^{-1}$) one can use the covariant 
formalism with $D = d+1$ oscillators to find,after replacing a 
discrete sum over $n$ by an integral over $\omega$,
\begin{equation}
\Gamma = c'_d \, \frac{G_N}{M \, \ell_s^2} \int d \omega \, 
\omega^{d-2} \left( \frac{n}{e^{\beta n} - 1} \right)^2 \, , 
\label{eq3.27}
\end{equation}
where $c'_d$ is a numerical constant, and where $n = M \, \ell_s^2 
\,\omega / 2$. The spectral decomposition of the total power 
radiated by the $\beta$-ensemble of strings is then simply deduced 
from (\ref{eq3.27}) by adding a factor $\hbar \, \omega$ in the 
integrand:
\begin{equation}
P = c'_d \, \frac{G_N}{M \, \ell_s^2} \int d \omega \, \omega^{d-1}
\left( \frac{n}{e^{\beta n} - 1} \right)^2 \, . \label{eq3.28}
\end{equation}

The results (\ref{eq3.27}), (\ref{eq3.28}) agree with corresponding
results (for closed strings) in the second reference \cite{halyo} 
and in \cite{amati} (note, however, that the factor $M^2$ in the 
equation (3.2) of \cite{amati} should be $M$ and that the constant 
contains $G_N$ and powers of $\ell_s$). The integrals 
(\ref{eq3.27}), (\ref{eq3.28}) are dominated by $n \sim \beta^{-1}$, 
i.e. $\omega \sim M_s$. This gives for the integrated quantities:
\begin{equation}
\Gamma \sim g^2 \, M \ , \ P \sim g^2 \, M \, M_s \, . 
\label{eq3.29}
\end{equation}
The second equation (\ref{eq3.29}) means that the mass of a highly
excited string decays exponentially, with half-evaporation time
\begin{equation}
\tau_{\rm evap}^{\rm string} \equiv M/P \sim \frac{\ell_s}{g^2} \, . 
\label{eq3.30}
\end{equation}
Let us anticipate on the next section and note that, at the 
transition $\lambda \equiv g^2 \, M / M_s \sim 1$ between string 
states and black hole states, not only the mass and the entropy are 
(in order of magnitude at least) continuous, but also the various 
radiative quantities: total luminosity $P$, half-evaporation time 
$\tau_{\rm evap}$, and peak of emission spectrum. Indeed, for a 
black hole decaying under Hawking radiation the temperature is 
$T_{\rm BH} \sim R_{\rm BH}^{-1}$ and
\begin{equation}
P_{\rm BH} \sim R_{\rm BH}^{-2} \sim \ell_s^{-2} \, \lambda^{-2 / 
(d-2)} \ , \ \tau_{\rm evap}^{\rm BH} \sim R_{\rm BH} \, S_{\rm BH} 
\sim \ell_s \, g^{-2} \, \lambda^{d/(d-2)} \, . \label{eq3.31}
\end{equation}

\section{Entropy of self-gravitating strings}
In the present section we shall combine the main results of the 
previous sections, Eqs.~(\ref{eq2.30}) and (\ref{eq3.25}), and 
heuristically extend them at the limit of their domain of validity. 
We consider a narrow band of string states that we follow when 
increasing adiabatically the string coupling $g$, starting from $g = 
0$\footnote{Alternatively, we can  consider the coupling as an
adjustable parameter (it is so in  perturbation theory) and just 
follow how different physical quantities change as $g$ is varied, 
whithout pretending that the change takes place in physical time.}. 
Let $M_0$, $R_0$ denote the ``bare'' values (i.e. for $g \rightarrow 
0$) of the mass and size of this band of states. Under the adiabatic 
variation of $g$, the mass and size, $M$, $R$, of this band of 
states will vary. However, the entropy$S(M,R)$ remains constant 
under this adiabatic process: $S(M,R) = S (M_0 , R_0)$. We assume, 
as usual, that the variation of $g$ is sufficiently slow to be 
reversible, but sufficiently fast to be able to neglect the decay of 
the states. We consider states with sizes $\ell_s \ll R_0 \ll M_0$ 
for which the correction factor,
\begin{equation}
c \, (R_0) \simeq (1 - c_1 \, R_0^{-2}) \, (1 - c_2 \, R_0^2 / 
M_0^2) \,,
\label{eq4.1}
\end{equation}
in the entropy
\begin{equation}
S (M_0 , R_0) = c \, (R_0) \, a_0 \, M_0 \, , \label{eq4.2}
\end{equation}
is near unity. [We use Eq.~(\ref{eq2.30}) in the limit $g 
\rightarrow 0$, for which it was derived.] Because of this reduced 
sensitivity of $c \,(R_0)$ on a possible direct effect of $g$ on $R$ 
(i.e. $R(g) = R_0 + \delta_g \, R$), the main effect of self-gravity 
on the entropy (considered as a function of the actual values $M$, 
$R$ when $g \not= 0$) will come from replacing $M_0$ as a function 
of $M$ and $R$. The mass-shift result (\ref{eq3.25}) gives $\delta 
\, M = M - M_0$ to first order in $g^2$. To the same 
accuracy\footnote{Actually, Eq.~(\ref{eq4.3}) is probably a more 
accurate version of the mass-shift formula because it
exhibits the real mass $M$ (rather than the bare mass $M_0$) as the
source of self-gravity.}, (\ref{eq3.25}) gives $M_0$ as a function 
of $M$ and $R$:
\begin{equation}
M_0 \simeq M + c_3 \, g^2 \, \frac{M^2}{R^{d-2}} = M \left( 1 + c_3 
\,\frac{g^2 \, M}{R^{d-2}} \right) \, , \label{eq4.3}
\end{equation}
where $c_3$ is a positive numerical constant.

Finally, combining Eqs.~(\ref{eq4.1})--(\ref{eq4.3}) (and 
neglecting, as just said, a small effect linked to $\delta_g \, R 
\not= 0$) leads to the following relation between the entropy, the 
mass and the size (all considered for self-gravitating states, with 
$g \not= 0$)
\begin{equation}
S(M,R) \simeq a_0 \, M \left( 1 - \frac{1}{R^2} \right) \left( 1 -
\frac{R^2}{M^2} \right) \left( 1 + \frac{g^2 \, M}{R^{d-2}} \right) 
\, . \label{eq4.4}
\end{equation}
For notational simplicity, we henceforth set to unity the 
coefficients $c_1$, $c_2$ and $c_3$. There is no loss of generality 
in doing so because we can redefine $\ell_s$, $R$ and $g$ to that 
effect, and use the corresponding (new) string units. The main point 
of the present paper is to emphasize that, for a given value of the 
total energy $M$ (and for some fixed value of $g$), the entropy 
$S(M,R)$ has a non trivial dependence on the radius $R$ of the 
considered string state. Eq.~(\ref{eq4.3}) exhibits two effects 
varying in opposite directions: (i) self-gravity favors small
values of $R$ (because they correspond to larger values of $M_0$, 
i.e. of the ``bare'' entropy), and (ii) the constraint of being of 
some fixed size $R$ disfavors both small $(R \ll \sqrt{M})$ and 
large $(R \gg \sqrt{M})$ values of $R$. For given values of $M$ and 
$g$, the most numerous (and therefore most probable) string states 
will have a size$R_* (M;g)$ which maximizes the entropy $S(M,R)$. 
Said differently, the total degeneracy of the complete ensemble of 
self-gravitating string states with total energy $M$ (and no {\it a 
priori} size restriction) will be given by an integral (where 
$\Delta \, R$ is the rms fluctuation of $R$ given by 
Eq.~(\ref{eq2.17}))
\begin{equation}
{\cal D} (M) \sim \int \frac{d R}{\Delta \, R} \, e^{S(M,R)} \sim 
e^{S(M,R_*)} \label{eq4.5}
\end{equation}
which will be dominated by the saddle point $R_*$ which maximizes 
the exponent.

The value of the most probable size $R_*$ is a function of $M$, $g$ 
and the space dimension $d$. To better see the dependence on $d$, 
let us first consider the case (which we generically assume) where 
the correction factors in Eq.~(\ref{eq4.4}) (parentheses on the 
right-hand-side) are very close to unity so that
\begin{equation}
S(M,R) \simeq a_0 \, M \, (1 - V(R)) \, , \label{eq4.6}
\end{equation}
where
\begin{equation}
V(R) = \frac{1}{R^2} + \frac{R^2}{M^2} - \frac{g^2 \, M}{R^{d-2}} \, 
.\label{eq4.7}
\end{equation}

One can think of $V(R)$ as an effective potential for $R$. The most
probable size $R_*$ must minimize $V(R)$. This effective potential 
can be thought of as the superposition of: (i) a centrifugal barrier 
near $R=0$ (coming from the result (\ref{eq2.29})), (ii) an harmonic 
potential (forbidding the large values of $R$), and (iii) an 
attractive (gravitational) potential. When $g^2$ is small the 
minimum of $V(R)$ will come from the competition between the 
centrifugal barrier and the harmonic potential and will be located 
around the value $R_*^{-2} \simeq R_*^2 / M^2$, i.e. $R_* \simeq 
\sqrt{M} = R_{\rm rw}$. This random walk value will remain (modulo 
small corrections) a local minimum of $V(R)$ (i.e. a local maximum 
of $S(M,R)$) as long as $g^2 \, M / R_*^{d-2} \ll R_*^{-2}$, i.e. 
for $g^2 \ll g_0^2$ with
\begin{equation}
g_0^2 \equiv M^{\frac{d-6}{2}} \, . \label{eq4.8}
\end{equation}
More precisely, working perturbatively in $g^2$, the minimization of 
$V(R)$ yields
\begin{equation}
R_* \simeq \sqrt{M} \left( 1 - \frac{d-2}{8} \, \frac{g^2}{g_0^2} 
\right) \, . \label{eq4.8bis}
\end{equation}

Note that, when $g^2 \ll g_0^2$, the value of $V(R)$ at this local
minimum is of order $V_{\rm min} \simeq + \, 2R_*^{-2} \simeq + \, 2
M^{-1}$, i.e. that it corresponds to a saddle-point entropy 
$S(M,R_*) \simeq a_0\, M (1 - V_{\rm min}) \simeq a_0 \, M - {\cal 
O} (1)$ which differs essentially negligibly from the ``bare'' 
entropy $a_0 \, M$ ($\gg 1$). To study what happens when $g^2$ 
further increases let us consider separately the various dimensions 
$d \geq 3$. We shall see that the special value $g_0^2$, 
Eq.~(\ref{eq4.8}), is significant (as marking a pre-transition, 
before the transition to the black hole state) only for $d=3$. For $d 
\geq 4$, the only special value of $g^2$ is the critical value
\begin{equation}
g_c^2 \sim M^{-1} \, , \label{eq4.9}
\end{equation}
around which takes place a transition toward a state more compact 
than the usual random walk one.

\subsection{$d=3$}
Let us first consider the (physical) case $d=3$, for which $g_0^2 
\sim M^{-3/2} \ll g_c^2 \sim M^{-1}$. In that case, when $g^2$ 
becomes larger than $g_0^2$, the (unique) local minimum of $V(R)$ 
slowly shifts towards values of $R$ lower than $R_{\rm rw}$ and 
determined by the competition between the centrifugal barrier 
$1/R^2$ and the gravitational potential$-g^2 \, M / R$.

In the approximation where we use the linearized form (\ref{eq4.6}),
(\ref{eq4.7}), and where (for $g^2 \gg g_0^2$) we neglect the term 
$R^2 / M^2$, the most probable size $R_*$ is
\begin{equation}
R_*^{(d=3)} \simeq \frac{2}{g^2 \, M} \ , \ \hbox{when} \ M^{-3/2} 
\ll g^2 \ll M^{-1} \, . \label{eq4.10}
\end{equation}

Note that as $g^2$ increases between $M^{-3/2}$ and $M^{-1}$, the 
most probable size $R_*^{(d=3)}$ smoothly interpolates between 
$R_{\rm rw}$ and a value of order unity, i.e. of order the string 
length. Note also that $V_{\rm min} \simeq -g^2 \, M / (2R_*)
\simeq -g^4 \, M^2 / 4$ remains smaller than one when $g^2 \lsim 
M^{-1}$ so that the saddle-point entropy $S(M,R_*) \simeq a_0 \, M 
(1 - V_{\rm min})$ never differs much from the ``bare'' value $a_0 
\, M$.

When $g^2$, in its increase, becomes comparable to $M^{-1}$, the 
radius becomes of order one and it is important to take into account 
the (supposedly) more exact expression (\ref{eq4.4}) (in which the 
factor $(1-R^{-2})$ plays the crucial role of cutting off any size 
$R \leq 1$). If we neglect, as above, the term $R^2 / M^2$ (which is 
indeed even more negligible in the region $R \sim 1$) but maximize 
the factored expression (\ref{eq4.4}), we find that the most 
probable size $R_*$ reads
\begin{equation}
R_*^{(d=3)} \simeq \frac{1 + \sqrt{1 + 3 \, \lambda^2}}{\lambda} 
\ , \ \hbox{when} \ g^2 \gg M^{-3/2} \, , \label{eq4.11}
\end{equation}
where we recall the definition
\begin{equation}
\lambda \equiv g^2 \, M \, . \label{eq4.12}
\end{equation}

When $\lambda \ll 1$, the result (\ref{eq4.11}) reproduces the 
simple linearized estimate (\ref{eq4.10}). When $\lambda \gsim 1$,
Eq.~(\ref{eq4.11}) says that the most probable size, when $g^2$
increases, tends  to a limiting size ($R_{\infty} = \sqrt 3$) 
slightly larger than the minimal one ($R_{\rm min} = 1$) 
corresponding to zero entropy. [Note that even for the formal 
asymptotic value $R_{\infty} = \sqrt 3$, the reduction in entropy 
due to the factor $1-R^{-2}$ is only $2/3$.] On the other hand, the 
fractional self-gravity $G_N \, M / R_*$ (which measures the 
gravitational deformation away from flat space), or the
corresponding term in Eq.~(\ref{eq4.4}), continues to increase with 
$g^2$ as
\begin{equation}
\frac{\lambda}{R_*} = \frac{\lambda^2}{1 + \sqrt{1+3 \, \lambda^2}} 
\, . \label{eq4.13}
\end{equation}
The right-hand side of Eq.~(\ref{eq4.13}) becomes unity for $\lambda 
=\sqrt 5 = 2.236$. The picture suggested by these results is that 
the string smoothly contracts, as $g$ increases, from its initial 
random walk size down to a limiting compact state of size slightly 
larger than $\ell_s$. For some value of $\lambda$ of order unity 
(may be between 1 and 2; indeed, even for $\lambda = 1$ the size 
$R_* = 2$ and the self-gravity $\lambda / R_* = 0.5$ suggest one may 
still trust a compact string description) the self-gravity of this 
compact string state will become so strong that one expects it to 
collapse to a black hole state. We recall that, as emphasized in 
Refs.~ \cite{Bowick}, \cite{susskind}, \cite{GVDivonne}, 
\cite{halyo}, \cite{hp1}, \cite{hp2}, when $\lambda \sim 1$, the 
mass  of the string state matches (in order of magnitude) that of a 
(Schwarzschild) black hole with Bekenstein-Hawking entropy equal to 
the string entropy $S$.

\subsection{$d=4$}

When $d=4$, the argument above Eq.~(\ref{eq4.8bis}) suggests that 
the random-walk size remains the most probable size up to $g^2 \lsim 
g_0^2\sim M^{-1}$, i.e. up to $\lambda \lsim 1$. A more accurate 
approximation to the most probable size $R_*$, when $\lambda < 1$, is 
obtained by minimizing exactly $V(R)$, Eq.~(\ref{eq4.7}). This 
yields
\begin{equation}
R_*^{(d=4)} \simeq M^{1/2} \, (1-\lambda)^{1/4} \ , 
\ \hbox{when} \ \lambda < 1 \, .
\label{eq4.14}
\end{equation}
This shows that the size will decrease, but one cannot trust this 
estimate when $\lambda \rightarrow 1^-$. To study more precisely 
what happens when $\lambda \sim 1$ we must take into
account the more exact factorized form (\ref{eq4.4}). Let us now 
neglect the $R^2 / M^2$ term and consider the approximation
\begin{equation}
S^{(d=4)} (M,R) \simeq a_0 \, M \left( 1 - \frac{1}{R^2} \right) 
\left(1 + \frac{\lambda}{R^2} \right) \, . \label{eq4.15}
\end{equation}
The right-hand side of Eq.~(\ref{eq4.14}) has a maximum only for 
$\lambda > 1$, in which case
\begin{equation}
R_*^{(d=4)} \simeq \left( \frac{2 \lambda}{\lambda - 1} 
\right)^{1/2}  \, \ \hbox{when} \ \lambda > 1 \, . \label{eq4.16}
\end{equation}
If we had taken into account the full expression (\ref{eq4.4}) the 
two results (\ref{eq4.14}), (\ref{eq4.16}), valid on each side of 
$\lambda = 1$, would have blended in a result showing that 
around\footnote{The transition takes place in the range $\vert 
\lambda - 1 \vert \sim M^{-2/3}$ corresponding to $R_* \sim 
M^{1/3}$.} $\lambda = 1$ the most probable size {\it continuously} 
interpolates between $R_{\rm rw}$ and a size of order $\ell_s$. Note 
that, according to Eq.~(\ref{eq4.16}), as $\lambda$ becomes $\gg 1$, 
$R_*^{(d=4)}$ tends to a limiting size ($R_{\infty} = \sqrt 2$) 
slightly larger than $R_{\rm min} = 1$ (corresponding to zero 
entropy). When $\lambda > 1$ the fractional self-gravity of the 
compact string states reads
\begin{equation}
\frac{\lambda}{R_*^2} = \frac{\lambda - 1}{2} \, . \label{eq4.17}
\end{equation}
As in the case $d=3$, one expects that for some value of $\lambda$
strictly larger than 1, the self-gravity of the compact string state 
will become so strong that it will collapse to a black hole state. 
Again the mass, size and entropy match (in order of magnitude) those 
of a black hole when $\lambda \sim 1$. The only difference between 
$d=4$ and $d=3$ is that the transition to the compact state, though 
still continuous, is sharply concentrated around $\lambda = 1$ 
instead of taking place over the extended range $M^{-1/2} \lsim 
\lambda \lsim 1$.

\subsection{$d \geq 5$}
When $d \geq 5$, the argument around Eq.~(\ref{eq4.8}) shows that 
the random walk size $R_{\rm rw} \simeq \sqrt{M}$ is a consistent
local maximum of the entropy in the whole domain $g^2 \ll g_0^2$, 
i.e. for $\lambda \equiv g^2 \, M \ll M^{\frac{d-4}{2}}$, which 
allows values $\lambda \gg 1$. However, a second, disconnected 
maximum of the entropy, as function of$R$, could exist. To 
investigate this we consider again (\ref{eq4.4}), when neglecting 
the $R^2 / M^2$ term (because we are interested in other
possible solutions with small sizes):
\begin{equation}
\frac{S(M,R)}{a_0 \, M} \simeq (1-x) \, (1+\lambda \, x^{\nu}) 
\equiv s(x) \, , \label{eq4.18}
\end{equation}
where we have defined $x \equiv R^{-2}$ and $\nu \equiv (d-2)/2$. By
studying analytically the maxima and inflection points of $s(x)$, 
one finds that, in the present case where $\nu = (d-2)/2 > 1$, there 
are two critical values $\lambda_1 < \lambda_2$ of the parameter 
$\lambda \equiv g^2 \,M$. The first one,
\begin{equation}
\lambda_1 = \left( \frac{\nu + 1}{\nu - 1} \right)^{\nu - 1} \ , \ 
R_1 = x_1^{-1/2} = \left( \frac{\nu + 1}{\nu - 1} \right)^{1/2} > 1 
\ , \ s_1 =1 - \left( \frac{\nu - 1}{\nu + 1} \right)^2 \, , 
\label{eq4.19}
\end{equation}
corresponds to the birth (through an inflection point) of a maximum 
and a minimum of the function $s(R)$. Because $s_1 < 1$ is strictly 
lower than the usual random walk maximum with $s (R_{\rm rw}) \simeq 
1 - 2/M \simeq 1$, the local maximum near $R \sim 1$ of the entropy, 
which starts to exist when $\lambda > \lambda_1$, is, at first, only 
metastable with respect to $R_{\rm rw}$. However, there is a second 
critical value of $\lambda$, $\lambda_2 > \lambda_1$, defined by
\begin{equation}
\lambda_2 = \nu \left( \frac{\nu}{\nu - 1} \right)^{\nu - 1} \ , \ 
R_2 = x_2^{-1/2} = \left( \frac{\nu}{\nu - 1} \right)^{\frac{1}{2}} 
> 1 \ , \ s_2 = 1 \, . \label{eq4.20}
\end{equation}
When $\lambda > \lambda_2$ the local maximum near $R \sim 1$ of the
entropy has $s(R) > 1$, i.e. it has become the global maximum of the 
entropy, making the usual random walk local maximum only metastable. 
Therefore,when $\lambda > \lambda_2$ the most probable string state 
is a very compact state of size comparable to $\ell_s$. Formally, 
this new global maximum exists for any $\lambda \gsim 1$ and tends, 
when $\lambda \rightarrow\infty$, toward the limiting location 
$R_{\infty} = ((\nu + 1) / \nu)^{1/2} > 1$, 
i.e. slightly (but finitely) 
above the minimum size $R =1$. However, as in the cases $d \leq 4$, 
the self-gravity of the stable compact string state will become 
strong when $\lambda \gsim 1$, so that it is expected to collapse 
(for some $\lambda_c > \lambda_2$) to a black hole state. As in the 
cases $d \leq 4$, the mass, size and entropy of this compact string 
state match those of a black hole. The big difference with
the cases $d \leq 4$ is that the transition between the (stable) 
random walk typical configuration and the (stable) compact one is 
discontinuous. Our present model suggests that (when $\nu \equiv 
(d-2) / 2 > 1$) a highly excited single string system can exist, 
when $\lambda > \lambda_2$, in two different stable typical states: 
(i) a dilute state of typical size $R_{\rm rw} \simeq \sqrt{M}$ and 
typical mean density $\rho \sim M / R_{\rm rw}^d \sim M^{-\nu} \ll 
1$, and (ii) a condensed state of typical size $R \sim1$ and typical 
mean density (using $\lambda \sim 1$): $\rho \sim M \sim g^{-2}
\gg 1$. We shall comment further below on the value $\rho \sim 
g^{-2}$ of the dense state of string matter.

\section{Discussion}
Technically, the main new result of the present work is the 
(dimension independent) estimate\footnote{In spite of our efforts in 
Section II, this result remains non rigorous and open to ${\cal O} 
(1/d)$ fractional corrections because of the difficulty to define a 
good quantum operator representing the mean radius of a string 
state.} $c(R) = 1 - c_1 / R^2$, with $c_1 \simeq (3/4) \, \ell_s^2 = 
3\alpha' / 2$, of the factor giving the decrease in the entropy 
$(2\pi ((d-1) \, \alpha' / 6)^{1/2} \, M)$ of a narrow band of very 
massive (open\footnote{For technical simplicity, we have restricted 
our attention to open bosonic strings. We could have dealt
with closed bosonic strings by doubling the oscillators, but the 
level matching condition, $N_L = N_R$, would have complicated the 
definition of the grand canonical ensemble we used. We expect that 
our results (whichare semi-classical) apply (with some numerical 
changes) to open or closed superstrings.}) string states $M \gg 
(\alpha')^{-1/2}$, when considering only string states of size $R$ 
(modulo some fractionally small grand-canonical-type fluctuations). 
We have also justified (by dealing explicitly with relativistic and 
quantum effects in a semi-classical approximation) and refined (by 
computing the numerical coefficient, Eq.~(\ref{eq3.26})) the naive 
estimate, $\delta M = -c_d \, G_N \, M^2 / R^{d-2}$, of the mass 
shift of a massive string state due to the exchange of long range 
fields (graviton, dilaton and axion). [The exchange of these
fields is expected to be the most important one both because very 
excited string states tend to be large, and because the 
corresponding interactions are attractive and cumulative with the 
mass.]

Conceptually, the main new result of this paper concerns the most
probable state of a very massive single\footnote{We consider states 
of a single string because, for large values of the mass, the 
single-string entropy approximates the total entropy up to 
subleading terms.} self-gravitating string.  By combining our 
estimates of the entropy reduction due to the size constraint, and 
of the mass shift we come up with the expression (\ref{eq4.4}) for 
the logarithm of the number of self-gravitating 
string states of size $R$.  Our analysis of the function $S(M,R)$ 
clarifies the correspondence \cite{susskind}, \cite{halyo}, 
\cite{hp1}, \cite{hp2} between string states and black holes.  In 
particular, our results confirm many of the results of \cite{hp2}, 
but make them (in our opinion) physically clearer by dealing 
directly with the size distribution, in real space, of an ensemble 
of string states.  When our results differ from those of \cite{hp2}, 
they do so in a way which simplifies the physical picture
and make even more compelling the existence of a correspondence 
between strings and black holes.  For instance, \cite{hp2} suggested 
that in $d=5$ there was a phenomenon of hysteresis, with a critical 
value $g_0^2 \sim M^{-1/2}$ for the string $\rightarrow$ black hole 
transition, and a different critical value $g_c^2 \sim M^{-1} \ll 
g_0^2$ for the inverse transition:  black hole $\rightarrow$ string. 
Also, \cite{hp2} suggested that in $d > 6$, most excited string 
states would never form black holes. The simple physical picture 
suggested\footnote{Our conclusions are not rigourously established 
because they rely on assuming the validity of the result 
(\ref{eq4.4}) beyond the domain ($R^{-2} \ll 1$, $g^2 \, M / R^{d-2}
\ll 1$) where it was derived. However, we find heuristically 
convincing to believe in the presence of a reduction factor of the 
type $1-R^{-2}$ down to sizes very near the string scale. Our 
heuristic dealing with self-gravity is less compelling because we do 
not have a clear signal of when strong gravitational field effects 
become essential.} by our results is the following: In any 
dimension, if we start with a massive string state and increase the 
string coupling $g$, a {\it typical} string state will,
eventually, become more compact and will end up, when $\lambda_c = 
g_c^2\, M \sim 1$, in a ``condensed state'' of size $R \sim 1$, and 
mass density $\rho \sim g_c^{-2}$. Note that the basic reason why 
small strings, $R \sim 1$, dominate in any dimension the entropy when 
$\lambda \sim 1$ is that they descend from string states with bare 
mass $M_0 \simeq M (1 + \lambda / R^{d-2}) \sim 2 M$ which are 
exponentially more numerous than less condensed string states 
corresponding to smaller bare masses.

The nature of the transition between the initial ``dilute'' state 
and the final ``condensed'' one depends on the value of
the space dimension $d$. [As explained below Eq.~(\ref{eq4.4}), we 
henceforth set to unity, by suitable redefinitions of $\ell_s$, $R$ and 
$g$, the coefficients $c_1$, $c_2$ and $c_3$.] In $d=3$, the transition 
is gradual: when
$\lambda < M^{-1/2}$ the size of a typical state is $R_*^{(d=3)} 
\simeq M^{1/2}(1-M^{1/2} \, \lambda / 8)$, when $\lambda >  M^{1/2}$ 
the typical size is $R_*^{(d=3)} \simeq (1 + (1+3 \, 
\lambda^2)^{1/2}) / \lambda$. In $d=4$, the transition toward a 
condensed state is still continuous, but most of the size evolution 
takes place very near $\lambda = 1$: when $\lambda <1$,
$R_*^{(d=4)} \simeq M^{1/2} (1-\lambda)^{1/4}$, and when $\lambda > 
1$,$R_*^{(d=4)} \simeq (2\lambda / (\lambda - 1))^{1/2}$, with some 
smooth blending between the two evolutions around $\vert \lambda - 1 
\vert \sim M^{-2/3}$. In $d \geq 5$, the transition is discontinuous 
(like a first order phase transition between, say, gas and liquid 
states). Barring the consideration of metastable (supercooled) 
states, on expects that when $\lambda = \lambda_2 \simeq \nu^{\nu} / 
(\nu - 1)^{\nu - 1}$ (with $\nu = (d-2) /2$), the most probable size 
of a string state will jump from $R_{\rm rw}$ (when $\lambda < 
\lambda_2$) to a size of order unity (when $\lambda > \lambda_2$).

Let us, for definiteness, write down in more detail what happens in 
$d=3$. After maximization over $R$, the entropy of a 
self-gravitating string is given, when $M^{-3/2} \ll g^2 \ll 
M^{-1}$, by
\begin{equation}
S(M) = S (M,R_* (M)) \simeq a_0 \, M \left[ 1 + \frac{1}{4} \, (g^2 
\, M)^2 \right] \, . \label{eq5.1}
\end{equation}
By differentiating $S$ with respect to $M$, one finds the 
temperature of the ensemble of highly excited single string states 
of mass $M$:
\begin{equation}
T \simeq T_{\rm Hag} \left( 1 - \frac{3}{4} \, (g^2 \, M)^2 \right) 
\, , \label{eq5.2}
\end{equation}
with $T_{\rm Hag} \equiv a_0^{-1}$. Eq.~(\ref{eq5.2}) explicitly 
exhibits the modification of the Hagedorn temperature due to 
self-gravity (in agreement with results of \cite{hp2} obtained by a 
completely different approach). Note that, both in 
Eqs.~(\ref{eq5.1}) and (\ref{eq5.2}), the self-gravity modifications 
are fractionally of order unity at the transition $g^2 \, M \sim 1$.

One can think of the ``condensed'' state of (single) string matter, 
reached (in any $d$) when $\lambda \sim 1$, as an analog of a 
neutron star with respect to an ordinary star (or a white dwarf). It 
is very compact (because of self gravity) but it is stable (in
some range for $g$) under gravitational collapse. However, if one 
further increases $g$ or $M$ (in fact, $\lambda = g^2 \, M$), the 
condensed string state is expected (when $\lambda$ reaches some 
$\lambda_3 > \lambda_2$, $\lambda_3 = {\cal O} (1)$) to collapse 
down to a black hole state (analogously to a neutron star collapsing 
to a black hole when its mass exceeds the Landau-Oppenheimer-Volkoff 
critical mass). Still in analogy with neutron stars, one notes that 
general relativistic strong gravitational field effects are crucial 
for determining the onset of gravitational collapse; indeed, under 
the ``Newtonian'' approximation (\ref{eq4.4}), the condensed string 
state could continue to exist for arbitrary large values of 
$\lambda$.

It is interesting to note that the value of the mass density at the
formation of the condensed string state is $\rho \sim g^{-2}$. This
is reminiscent of the prediction by Atick and Witten \cite{AW88} of 
a first-order phase transition of a self-gravitating thermal gas of 
strings, near the Hagedorn temperature\footnote{Note that, by 
definition, in our {\it single} string system, the formal 
temperature $T = (\partial S / \partial M)^{-1}$ is always near the 
Hagedorn temperature.}, towards a dense state with energy
density $\rho \sim g^{-2}$ (typical of a genus-zero contribution to 
the free energy). Ref.~\cite{AW88} suggested that this transition is 
first-order because of the coupling to the dilaton. This suggestion 
agrees with our finding of a discontinuous transition to the single 
string condensed state in dimensions $\geq 5$ (Ref.~\cite{AW88} work 
in higher dimensions, $d=25$ for the bosonic case). It would be 
interesting to deepen these links between self-gravitating single 
string states and multi-string states.

Assuming the existence (confirmed by the present work) of a dense 
state of self-gravitating string matter with energy density $\rho 
\sim g^{-2}$, it would be fascinating to be able to explore in 
detail (with appropriate, strong gravity tools) its gravitational 
dynamics, both in the present context of a single, isolated object 
(``collapse problem''), and in the cosmological context (problem of 
the origin of the expansion of the universe).

Let us come back to the consequences of the picture brought 
by the present work for the problem of the end point of the 
evaporation of a Schwarzschild black hole and the interpretation of 
black hole entropy. In that case one fixes the value of $g$ (assumed 
to be $\ll 1$) and considers a black hole which slowly looses its 
mass via Hawking radiation. When the mass gets as low as a 
value\footnote{Note that the mass at the black hole $\rightarrow$ 
string transition is larger than the Planck mass $M_P \sim 
(G_N)^{-1/2} \sim g^{-1}$ by a factor $g^{-1} \gg 1$.} $M \sim 
g^{-2}$, for which the radius of the black hole is of order one 
(in string units), one expects the black hole to transform (in all
dimensions) into a typical string\footnote{A check on the 
single-string dominance of the transition black hole $\rightarrow$ 
string is to note that the single string entropy $\sim M / M_s$ is 
much larger than the entropy of a ball of radiation $S_{\rm rad} 
\sim (RM)^{d/(d+1)}$ with size $R \sim R_{\rm BH} \sim \ell_s$ at 
the transition.} state corresponding to $\lambda = g^2 \, M \sim 1$, 
which is a dense state (still of radius $R \sim 1$). This 
string state will further decay and loose mass, predominantly via 
the emission of massless quanta, with a quasi thermal spectrum with 
temperature $T \sim T_{\rm Hagedorn} = a_0^{-1}$ (see 
Eq.~(\ref{eq3.28}) and Refs.~\cite{halyo}, 
\cite{amati}) which smoothly matches the previous black hole Hawking 
temperature. This mass loss will further decrease the product 
$\lambda = g^2 \, M$, and this decrease will, either gradually or 
suddenly, cause the initially compact string state to inflate to 
much larger sizes. For instance, if $d \geq 4$, the string state
will quickly inflate to a size $R \sim \sqrt{M}$. Later, with 
continued mass loss, the string size will slowly shrink again toward 
$R \sim 1$ until a remaining string of mass $M \sim 1$ finally 
decays into stable massless quanta. In this picture, the black hole 
entropy acquires a somewhat clear statistical significance (as the 
degeneracy of a corresponding typical string state) {\it only} when 
$M$ and $g$ are related by $g^2 \, M \sim 1$. If we allow ourselves 
to vary (in a Gedanken experiment) the value of $g$ this gives a 
potential statistical significance to any black hole entropy value 
$S_{\rm BH}$ (by choosing $g^2 \sim S_{\rm BH}^{-1}$). We do not 
claim, however, to have a clear idea of the direct statistical meaning 
of $S_{\rm BH}$ when $g^2 \, S_{\rm BH} \gg 1$. Neither do we 
clearly understand the fate of the very large space (which could be 
excited in many ways) which resides inside very large classical 
black holes of radius $R_{\rm BH}\sim (g^2 \, S_{\rm BH})^{1/(d-1)} 
\gg 1$. The fact that the interior of a black hole of given mass could 
be 
{\it arbitrarily} large\footnote{E.g., in the Oppenheimer-Snyder 
model, one can join an arbitrarily large closed Friedmann dust universe, 
with hyperspherical opening angle $0 \leq \chi_0 \leq \pi$ arbitrarily 
near 
$\pi$, onto an exterior Schwarzschild spacetime of given mass $M$.}, 
and therefore arbitrarily complex, suggests that black hole physics 
is not exhausted by the idea (confirmed in the present paper) of a 
reversible transition between string-length-size black holes and 
string states.

On the string side, we also do not clearly understand how one could 
follow in detail (in the present non BPS framework) the 
``transformation'' of a strongly self-gravitating string state into 
a black hole state.

Finally, let us note that we expect that self-gravity will lift 
nearly completely the degeneracy of string states. [The degeneracy 
linked to the rotational symmetry, i.e. $2J + 1$ in $d=3$, is 
probably the only one to remain, and it is negligible compared to 
the string entropy.] Therefore we expect that the separation $\delta 
\, E$ between subsequent (string and black hole) energy levels will 
be exponentially small: $\delta \, E \sim \Delta \, M \, \exp 
(-S(M))$, where $\Delta \, M$ is the canonical-ensemble fluctuation 
in $M$. Such a $\delta \, E$ is negligibly small compared to the 
radiative width $\Gamma \sim g^2 \, M$ of the levels. This seems to 
mean that the discreteness of the quantum levels of strongly 
self-gravitating strings and black holes is very much blurred, and 
difficult to see observationally.

\section*{Acknowledgements}
This work has been clarified by useful suggestions from M.~Douglas,
K.~Gawedzki, M.~Green, I.~Kogan, G.~Parisi, A.~Polyakov, and (last 
but not least) M.~Vergassola. We wish also to thank A.~Buonanno for 
collaboration at an early stage and D.~Gross, J.~Polchinski, 
A.~Schwarz, L.~Susskind  and A.~Vilenkin for discussions. T.D. 
thanks the Theory Division of CERN, Gravity Probe B (Stanford 
University), and the Institute for Theoretical Physics (Santa 
Barbara) for hospitality. Partial support from NASA grant NAS8-39225 
is acknowledged. G.V. thanks the IHES for hospitality during the 
early, crucial stages of this work.

\end{document}